\pdfoutput=1
\newcommand*{\ATLASLATEXPATH}{./}
\documentclass[UKenglish,texlive=2016,cernpreprint]{\ATLASLATEXPATH atlasdoc}
\usepackage{\ATLASLATEXPATH atlaspackage}
\usepackage{\ATLASLATEXPATH atlasbiblatex}

\usepackage{\ATLASLATEXPATH atlascontribute}

\usepackage{\ATLASLATEXPATH atlasphysics}
\graphicspath{{logos/}{./}}

\def\mtop{\ensuremath{m_{\textrm{top}}}}
\def\fbkgd{\ensuremath{F_{\textrm{bkgd}}}}
\def\rtt{\ensuremath{R_{\textrm{3/2}}}}
\def\nbtag{\ensuremath{N_{b_\textrm{tag}}}}

\AtlasTitle{Top-quark mass measurement in the all-hadronic $\ttbar$ decay channel at $\rts = \SI{8} \TeV$ with the ATLAS detector}

\author{The ATLAS Collaboration}

\PreprintIdNumber{CERN-EP-2016-264} 

\AtlasJournal{JHEP}
\AtlasAbstract{%
 The top-quark mass is measured in the
  all-hadronic top-antitop quark decay channel using
  proton--proton collisions at a centre-of-mass energy of 
  $\rts = 8\,\textrm{TeV}$ with the ATLAS detector at the CERN Large 
  Hadron Collider. The data set used in the analysis corresponds 
  to an integrated luminosity of $\SI{20.2}\ifb$.
  The large multi-jet background
  is modelled using a data-driven method.
  The top-quark mass is obtained from template fits
  to the ratio of the three-jet to the dijet mass.
  The three-jet mass is obtained from the three jets assigned to
  the top quark decay. From these three jets
  the dijet mass is obtained using the two jets assigned to the
  $\Wboson$ boson decay.
  The top-quark mass is measured to be
  $173.72 \pm 0.55 \stat \pm 1.01 \syst\,\textrm{GeV}$.
}

\hypersetup{pdftitle={ATLAS document},pdfauthor={The ATLAS Collaboration}}

\begin{document}

\maketitle

%\tableofcontents

% List of contributors - print here or after the Bibliography.
%\PrintAtlasContribute{0.30}
%\clearpage

%-------------------------------------------------------------------------------
\section{Introduction}
\label{sec:intro}
%-------------------------------------------------------------------------------

Of all known fundamental particles, the top quark has 
the largest mass.
Its existence was predicted in 1973 by Kobayashi and 
Maskawa~\cite{Koba}, and it was not observed directly 
until 1995, by the CDF and D0 experiments at the 
Tevatron~\cite{cdf,d0}. Since 2010, top quarks have 
also been observed at the Large Hadron Collider 
(LHC)~\cite{Evans:2008zzb} at CERN. 
Due to the higher centre-of-mass energy, top quark 
production at the LHC is an order of magnitude larger 
than at the Tevatron. The large data sets of top--antitop 
quark ($\ttbar$) pairs allow many precision studies and 
measurements of top quark properties. The Yukawa coupling 
of the top quark is predicted to be close to 
unity~\cite{PhysRevD.8.1226,Weinberg:1990me}, suggesting 
that it may play a special role in electroweak symmetry 
breaking. 
In the Standard Model (SM), the top quark dominantly 
contributes to the quantum corrections to the Higgs 
self coupling~\cite{Hollik:1995dv,Peskin:2015kka}. 
Precise measurements of the top-quark mass ($\mtop$) 
are therefore very important in probing the stability of 
the vacuum~\cite{Degrassi:2012ry,Baak:2014ora}, and contribute 
to searches for signs of physics beyond the SM.

Today the most precise individual measurement of $\mtop$ is 
in the single-lepton decay channel of top--antitop quark pairs, 
where one top quark decays into a $b$-quark, a charged lepton 
and a neutrino and the other top quark decays into a $b$-quark 
and two $u$/$d$/$c$/$s$-quarks, performed by the CMS Collaboration, 
yielding a value of 
$\mtop = 172.35 \pm 0.16 \stat \pm 0.48 \syst\,\textrm{GeV}$~\cite{Khachatryan:2015hba}.
The most precise measurement of $\mtop$ in the dileptonic 
$\ttbar$ decay channel, where each of the top quarks decays 
into a $b$-quark, a charged lepton and its neutrino, is from 
the ATLAS Collaboration, yielding a value of 
$\mtop = 172.99 \pm 0.41 \stat \pm 0.74 \syst\,\textrm{GeV}$~\cite{Aaboud:2016igd}.
Further $\mtop$ results are available in 
\mbox{Refs.~\cite{ATLAS:2014wva,PhysRevLett.113.032002,Aad:2015nba}}.

The top-quark mass measurement in the all-hadronic $\ttbar$  
channel takes advantage of the largest branching ratio (46\%) 
among the possible top quark decay channels~\cite{Agashe:2014kda}.
The all-hadronic channel involves six jets at leading order, 
two originating from $b$-quarks and four originating from 
the two $\Wboson$ boson hadronic decays.
It is a challenging measurement because of the large multi-jet 
background arising from various quantum chromodynamics (QCD) 
processes, which can exceed the $\ttbar$ production by several 
orders of magnitude. However, all-hadronic $\ttbar$ events 
profit from having no neutrinos among the decay products, 
so that all four-momenta can be measured directly.
The multi-jet background for the all-hadronic $\ttbar$ 
channel, while large, leads to different systematic uncertainties 
than in the case of the single- and dileptonic $\ttbar$ channels. 
Thus, all-hadronic analyses offer an opportunity to cross-check 
top-quark mass measurements performed in the other channels.
The most recent measurements of $\mtop$ in the all-hadronic 
channel were performed by the CMS Collaboration with
$\mtop =$ 172.32 $\pm$ 0.25 $\stat \pm 0.59 \syst\,\textrm{GeV}$~\cite{Khachatryan:2015hba}, 
and the ATLAS Collaboration with
$\mtop = 175.1 \pm 1.4 \stat \pm 1.2 \syst\,\textrm{GeV}$~\cite{Aad:2014zea}.

This paper presents a top-quark mass measurement in the 
$\ttbar$ all-hadronic channel using data collected by the ATLAS 
experiment in 2012.
The $\mtop$ measurement is obtained from template fits to the 
distribution of the ratio of three-jet to dijet masses 
($\rtt = m_{jjj}/m_{jj}$), similarly to a previous measurement 
at $\rts = 7\,\textrm{TeV}$~\cite{Aad:2014zea}. The three-jet mass 
is obtained from the three jets assigned to the top quark decay.
From the selected three jets the dijet mass is obtained using 
the two jets assigned to the $\Wboson$ boson decay.
The jet assignment is accomplished by using a $\chi^2$~fit to 
the $\ttbar$~system, so there are two values of $\rtt$ measured 
in each event. The observable $\rtt$  employed in this analysis 
achieves a partial cancellation of systematic effects common to 
the masses of the reconstructed top quark and associated $\Wboson$ 
boson, notably the significant uncertainty on the jet energy 
scale. Data-driven techniques are used to estimate the contribution 
from multi-jet background events. 
Data events are divided into several disjoint regions using 
two uncorrelated observables. The region containing the largest 
relative fraction of $\ttbar$ events is labeled the signal region. 
The background is estimated from the other regions, which determine 
the shape of the background distribution in the signal region. 

The paper is organised as follows. After a brief description of 
the ATLAS detector in Section~\ref{sec:ATLAS}, the data and Monte 
Carlo (MC) samples used in the analysis are described in 
Section~\ref{sec:DataMC}. The analysis event selection is further 
detailed in Section~\ref{sec:selection}. 
Section~\ref{sec:Top reconstruction} describes the method used 
to select the candidate four-momenta that comprise the 
reconstructed $\ttbar$ system. 
The estimation of the multi-jet background is detailed in 
Section~\ref{sec:Multijet background}. 
The method used to measure the top-quark mass and its uncertainties 
are reported in Sections~\ref{sec:Top-quark measurements}, 
~\ref{sec:Closure}, and ~\ref{sec:systematic uncertainties}.
The results of the measurement are presented in 
Section~\ref{sec:top measurement}, and the analysis is summarised 
in Section~\ref{sec:results}.

%-------------------------------------------------------------------------------
\section{ATLAS detector}
\label{sec:ATLAS}
%-------------------------------------------------------------------------------
%
The ATLAS detector~\cite{atlasdet:2008} is a multi-purpose particle
physics experiment with a forward-backward symmetric cylindrical
geometry and near $4 \pi$ coverage in solid angle~\footnote{The coordinate 
system used to describe the ATLAS detector is briefly summarised here. 
The nominal interaction point is defined as the origin of the coordinate 
system, while the beam direction defines the $z$--axis and the $x$--$y$ plane 
is transverse to the beam direction. The positive $x$--axis is defined 
as pointing from the interaction point to the centre of the
LHC ring and the positive $y$--axis is defined as pointing upwards. 
The azimuthal angle $\phi$ is measured around the beam axis, and the 
polar angle $\theta$ is the angle from the beam axis. The pseudorapidity 
is defined as $\eta = -\ln\tan(\theta/2)$.
The transverse momentum $\pt$, the transverse energy $\et$,
and the missing transverse momentum ($\met{}$) are defined in the
$x$--$y$ plane unless stated otherwise. The distance $\Delta R$
in the $\eta$--$\phi$ angle space is defined as
$\Delta R = \sqrt{(\Delta \eta)^2 + (\Delta \phi)^2}$.}. 
The inner tracking detector (ID) covers the pseudorapidity 
range $|\eta| < 2.5$, and consists of a silicon pixel detector, a silicon
microstrip detector, and, for $|\eta| < 2.0$, a transition
radiation tracker. The ID is surrounded by a thin
superconducting solenoid providing a $2 \,\textrm{T}$ magnetic field. A
high-granularity lead/liquid-argon (LAr) sampling electromagnetic
calorimeter covers the region $|\eta| < 3.2$. A
steel/scintillator-tile calorimeter provides hadronic coverage in the
range $|\eta| < 1.7$. LAr technology is also used for the hadronic
calorimeters in the endcap region $1.5 < |\eta| < 3.2$ and for
electromagnetic and hadronic measurements in the forward region 
up to $|\eta| = 4.9$. The muon spectrometer surrounds the calorimeters.  
It consists of three large air-core superconducting toroid systems, precision 
tracking chambers providing accurate muon tracking for $|\eta|<2.7$, and 
additional detectors for triggering in the region $|\eta| < 2.4$.

%------------------------------------------------------------------------------
\section{Data and Monte Carlo simulation}
\label{sec:DataMC}
%------------------------------------------------------------------------------
%
This analysis is performed using the proton--proton ($pp$)
collision data set at a centre-of-mass energy of 
$\rts = 8\,\textrm{TeV}$ collected with the ATLAS detector at 
the LHC. The data correspond to an integrated luminosity 
of $\SI{20.2}\ifb$. 
Samples of simulated MC events are used to optimise 
the analysis, to study the detector response and the 
efficiency to reconstruct $\ttbar$ events, to build signal 
template distributions used for fitting the top-quark mass, 
and to estimate systematic uncertainties.
Most of the MC samples used in the analysis are based on 
a full simulation of the ATLAS detector~\cite{Aad:2010ah} 
obtained using GEANT4~\cite{Agostinelli:2002hh}.
Some of the systematic uncertainties are studied using 
alternative $\ttbar$ samples processed through a faster 
ATLAS simulation (AFII) using parameterised showers in 
the calorimeters~\cite{ATLAS:1300517}. 
Additional simulated $pp$ collisions generated with 
\textsc{Pythia}~\cite{Sjostrand:2007gs} are overlaid to 
model the effects of additional collisions in the same 
and nearby bunch crossings (pile-up). All simulated events 
are processed using the same reconstruction algorithms 
and analysis chain as used for the data.

The nominal $\ttbar$ simulation sample is generated 
using the next-to-leading-order (NLO) MC program 
\textsc{POWHEG-BOX}~\cite{Nason:2004rx,Frixione:2007vw,Alioli:2010xd} 
with the NLO parton distribution function (PDF) set 
\textsc{CT10}~\cite{Lai:2010vv,Gao:2013xoa}, interfaced to 
\textsc{Pythia  6.427}~\cite{Sjostrand:2006za} with 
a set of tuned parameters called the \textsc{Perugia 2012} 
tune~\cite{Skands:2010ak} for parton shower, fragmentation 
and underlying-event modelling.
For the construction of the signal templates, MC events 
are generated at five different assumed values of $\mtop$, 
between $167.5$ and $177.5\,\textrm{GeV}$, in steps of 
$2.5\,\textrm{GeV}$.
The full simulation of the ATLAS detector sample at 
$172.5\,\textrm{GeV}$ has the largest number of generated 
events, and is used as the nominal signal sample. 
The $h_{\textrm{damp}}$ parameter~\cite{ATL-PHYS-PUB-2014-005}, 
which regulates the high-$\pt$ radiation in \textsc{POWHEG-BOX}, 
is set to the same $\mtop$ value as used in each of the 
generated \textsc{POWHEG-BOX} samples.
All the simulated samples used to estimate systematic 
uncertainties are further described in 
Section~\ref{sec:systematic uncertainties}.

All MC samples are normalised using the predicted top--antitop 
quark pair cross-section ($\sigma_{\ttbar}$) at 
\mbox{$\rts = 8\,\textrm{TeV}$}. For $\mtop = 172.5\,\textrm{GeV}$, 
the next-to-next-to-leading-order cross-section of
$\sigma_{\ttbar} = 253^{+13}_{-15}$ pb is calculated using 
the program \textsc{Top++2.0}~\cite{Czakon:2011xx}, which 
includes re-summation of next-to-next-to-leading logarithmic 
soft gluon terms.

%-------------------------------------------------------------------------------
\section{Event selection}
\label{sec:selection}
%-------------------------------------------------------------------------------

Events in this analysis are selected by a trigger that 
requires at least five jets with $\pt > 55\,\textrm{GeV}$. 
Only events with a well-reconstructed primary vertex 
formed by at least five tracks with $\pt > 400\,\textrm{MeV}$ 
are considered for the analysis. 
Events with isolated electrons (muons) with 
$E_{\textrm{T}} > 25\,\textrm{GeV}$ ($\pt> 20\,\textrm{GeV}$) and 
reconstructed in the central region of the detector 
within $|\eta| < 2.5$ are rejected.
Both lepton types are identified using the tight
working points as specified in Refs.~\cite{Aad:2011mk,Aad:2014rra}.
Jets ($j$) are reconstructed using the anti-${\mathit k_{\textrm {t}}}$ 
algorithm with radius parameter $R = 0.4$~\cite{antiktjets:2008}
employing topological clusters~\cite{caloclusref} in the 
calorimeter. These jets are calibrated to the hadronic 
energy scale as described in 
Refs.~\cite{Aad:2014bia,Aad:2011he,Aad:2015ina}.
The four-vector of the highest-energy muon ($\mu$) from 
among those matched within $\Delta R(j,\mu) < 0.3$ to 
a reconstructed jet, is added to the reconstructed jet 
four-vector. This is done to compensate for the energy 
losses in the calorimeter arising from semimuonic quark 
decays. In simulation this correction slightly improves 
both the jet energy response and resolution across the 
full range of jet energies. 

To ensure that the selected events are in the plateau 
region of the trigger efficiency curve where the trigger 
efficiency in data is greater than $90$\%, at least five 
of the reconstructed central jets (within $|\eta|< 2.5$) 
are required to have $\pt > 60\,\textrm{GeV}$. 
Any additional jet is required to have $\pt > 25\,\textrm{GeV}$ 
and $|\eta| < 2.5$.
All selected jets in an event must be isolated; any 
pairing of two jets ($j_i$ and $j_k$) reconstructed 
with the above criteria are required to not overlap 
within $\Delta\, R(j_i,j_k) < 0.6$. 
Events with jets failing this isolation requirement 
are rejected.

Events containing neutrinos are removed by requiring 
$\met{} < 60\,\textrm{GeV}$.
The $\met{}$ in an event is computed as the sum of a 
number of different terms~\cite{Aad:2016nrq,ATLAS-CONF-2014-019}.
Muons, electrons and jets are accounted for using the 
appropriate calibrations for each object. For each 
term considered, the missing transverse momentum is 
calculated as the negative sum of the calibrated 
reconstructed objects, projected onto the $x$ and 
$y$ directions.

For the final selection, events are kept if at least 
two of the six leading transverse momentum jets are 
identified as originating from a $b$-quark. 
Such jets are said to be $b$-tagged.
A neural network trained on decay vertex 
properties~\cite{ATLAS-CONF-2011-102} is used 
to identify these $b$-tagged jets. 
Because of the large number of $c$-quarks 
originating from the $\Wboson$ boson decays in 
this analysis (on average one $c$-quark per 
$\ttbar$ event) a $b$-tagger trained to reject 
$u$/$d$/$s$-jets but also a large fraction of 
$c$-jets  is used. 
Events with fewer than two $b$-tagged jets are 
used for the background estimate described in 
Section~\ref{sec:Multijet background}.
The chosen working point for the $b$-tagging 
neural network has an identification efficiency 
of about $57$\% ~\cite{ATLAS-CONF-2014-046} for 
jets from $b$-quarks, with a rejection factor of 
about $330$ for jets arising from $u$/$d$/$s$-quarks, 
and a factor of about $13$ for jets arising from 
$c$-quarks.

In each event the two jets with leading $b$-tag 
weights ($b_i$ and $b_j$) are required to 
satisfy $\Delta \phi(b_i,b_j) > 1.5$.
The quantities $b_i$ and $b_j$ represent here 
the vectorial transverse momentum of a $b$-jet: 
$\boldmath{b}_{\mathrm{T},i}$ and $\boldmath{b}_{\mathrm{T},j}$.
This $\Delta \phi$ cut is very powerful in rejecting 
combinatorial background events; most of these are 
true $\ttbar$ events where the incorrect jets are 
associated with the top quark.
Finally, a cut is applied based on the azimuthal 
angle between $b$-jets and their associated $\Wboson$ 
boson candidate: the average of the two angular 
separations for each event is required to satisfy
$\langle\Delta\phi(b,\Wboson)\rangle < 2$. Here the 
$b$, and the $\Wboson$ are the vectorial transverse 
momentum of a $b$-jet and a $\Wboson$ boson: 
$\boldmath{b}_{\mathrm{T}}$ and $\boldmath{W}_{\mathrm{T}}$,
identified by means of the three-jet combination 
that best fits the $\ttbar$ event hypothesis described 
in Section~\ref{sec:Top reconstruction}.
This $\Delta\phi$ cut rejects a large fraction of events 
from the multi-jet and combinatorial backgrounds, 
as well as events from non-all-hadronic $\ttbar$ decays.
Events failing this final selection cut are, however 
retained for the purpose of modelling the multi-jet 
background, as detailed in Section~\ref{sec:Multijet background}.

Table~\ref{tab:selection} summarises the yields obtained 
after each of the individual selection cuts.
The $\chi^2$ cut listed in Table~\ref{tab:selection} is 
described in Section~\ref{sec:Top reconstruction}. The 
number of $b$-tagged jets ($\nbtag$) and 
$\langle\Delta\phi(b,\Wboson)\rangle$ are the two 
observables used for the data-driven multi-jet background 
estimation, further detailed in Section~\ref{sec:Multijet background}. 

\begin{table}{}
\begin{center}
\begin{tabular}{l||r|rcl}
\cline{2-5}
\multicolumn{1}{c}{} & \multicolumn{4}{c}{ Event yields (thousands)}                \\ 
\cline{2-5}
Cut & \multicolumn{1}{ r| }{Data } & \multicolumn{3}{r}{$\ttbar$ all-hadronic (MC)} \\ 
\hline 
\hline
 Initial                              & $850450$  & $2338$  & $\pm$ & $1$           \\ 
$N_{{\textrm{PV}}_{>4 \,\textrm{tracks}}}$ \& no isolated $e$/$\mu$ & $33476$  & $308.7$ & $\pm$ & $0.6$\\
 Trigger: $5$ jets with $\pt > 55\,{\textrm{GeV}}$ \& $\ge 6$ good jets & $16110$ & $241.4$ & $\pm$ & $0.5$  \\ 
 No $2$ good jets ($j_i,j_k$) within $\Delta R(j_i,j_k) < 0.6$ & $7646$   & $142.9$ & $\pm$ & $0.4$     \\ 
 $\ge 5$ good jets with $\pt > 60\,{\textrm{GeV}}$  & $3303$   & $51.4$  & $\pm$ & $0.2$ \\ 
 $\met{} < 60\,{\textrm{GeV}}$             & $3021$   & $46.3$  & $\pm$ & $0.2$          \\ 
 $\Delta \phi$($b_i,b_j$) $> 1.5$     & $1737$   & $30.9$  & $\pm$ & $0.2$          \\ 
 $\chi^2 < 11$                        & $645.8$  & $22.3$  & $\pm$ & $0.1$          \\ 
 $\nbtag \geq 2$                      & $21.9$   & $6.61$  & $\pm$ & $0.08$         \\ 
 $\langle\Delta\phi(b,\Wboson)\rangle < 2$  & $12.9$   & $4.40$  & $\pm$ & $0.07$   \\ 
\hline
\hline
\end{tabular}
  \caption{Event yields following each of the individual event 
selection cuts, with values shown for both the data and 
all-hadronic MC events generated at $\mtop = 172.5\,\textrm{GeV}$ 
(shown with statistical uncertainty). 
The $\ttbar$ contribution is after scaling to the 
theoretical cross-section and integrated luminosity. 
$N_{\textrm{PV}_{>4 \,\textrm{tracks}}}$ is the number 
of primary vertices with $> 4$ tracks. Good jets have 
$\pt > 25 \,\textrm{GeV}$ and $|\eta| < 2.5$.}
  \label{tab:selection}
  \end{center}
\end{table}

%-------------------------------------------------------------------------------
\section{$\ttbar$ reconstruction}
\label{sec:Top reconstruction}
%-------------------------------------------------------------------------------
%
In each event the $\ttbar$ final state is reconstructed
using all the jets from the all-hadronic $\ttbar$ decay chain: 
$\ttbar\to b\Wboson b \Wboson \to b_1 j_1 j_2 \, b_2 j_3 j_4$. 
To determine the top-quark mass in each $\ttbar$ event, 
a minimum-$\chi^2$ approach is adopted, with the $\chi^2$ 
defined as:

\begin{equation} 
   \chi^2 = \frac{(m_{b_1 j_1 j_2} - m_{b_2 j_3 j_4})^{2}}{\sigma_{\Delta m_{bjj}}^{2}} + 
                    \frac{(m_{j_1 j_2} - m_{\Wboson}^{\textrm{MC}})^{2}}{\sigma_{m_{\Wboson}^{\textrm{MC}}}^{2}} +
                    \frac{(m_{j_{3} j_{4}} - m_{\Wboson}^{\textrm{MC}})^{2}}{\sigma_{m_{\Wboson}^{\textrm{MC}}}^{2}}. 
 \label{eq:newchi2}
\end{equation}

Here, two of the reconstructed jets are associated with the 
bottom-type quarks produced directly from the top quark and 
antitop quark decays ($b_1$ and $b_2$), the other four jets 
are assumed to be $u$/$d$/$c$/$s$-quark jets from the 
$\Wboson$ boson hadronic decay ($j_i$, where $i= 1, \dots, 4$), 
and $\Delta m_{bjj} = m_{b_1 j_1 j_2} - m_{b_2 j_3 j_4}$.
This method considers all possible permutations of the six 
or more reconstructed jets in each event. The permutation 
resulting in the lowest $\chi^2$ value is kept. A low 
$\chi^2$ value indicates a permutation of jets consistent 
with the $\ttbar$ hypothesis.
No explicit $b$-tagging information is used in Eq.~(\ref{eq:newchi2}).

In each combination the reconstructed masses of the two 
hadronically decaying $\Wboson$ bosons ($m_{j_1\,j_{2}}$ 
and $m_{j_3\, j_4}$) in data are compared to the mean of 
the mass distribution of correctly reconstructed $\Wboson$ 
bosons in simulated signal MC events ($m_{\Wboson}^{\textrm{MC}}$).
The correct reconstruction of the top quarks and the 
$\Wboson$ bosons in a simulated event is achieved by 
matching parton-level particles to the event's jets.
The widths ($\sigma_{m_{W}^{\textrm{MC}}}$ and 
$\sigma_{\Delta m_{bjj}}$) used in the denominators of 
Eq.~(\ref{eq:newchi2}) are obtained from fits to a single 
Gaussian function to the mass distributions of the correctly 
reconstructed top quarks and $\Wboson$ bosons:
$\sigma_{\Delta m_{bjj}} = 21.60 \pm 0.16 \stat\,{\textrm{GeV}}$ and 
$\sigma_{m_{W}^{\textrm{MC}}} = 7.89 \pm 0.05 \stat\,\textrm{GeV}$. 
The $m_{W}^{\textrm{MC}}$ mean value used in Eq.~(\ref{eq:newchi2}) 
is determined to be $81.18 \pm 0.04 \stat\,\textrm{GeV}$. 
To reduce the multi-jet background in the analysis and to 
eliminate events where the top quarks and the $\Wboson$ 
bosons in an event are not reconstructed correctly, a minimum
$\chi^2 < 11$ is required.

%-------------------------------------------------------------------------------
\section{Multi-jet background estimation}
\label{sec:Multijet background}
%-------------------------------------------------------------------------------
%
The available MC generators for multi-jet production include 
only leading-order theory calculations for final states with 
up to six partons. Therefore, the dominant multi-jet background 
in this analysis is determined directly from the data. 
Two largely uncorrelated variables are used to divide the data 
events into four different regions, such that the background 
is determined in the control regions and extrapolated to the 
signal region. The two chosen observables are the $\nbtag$ 
in an event, and the $\langle\Delta\phi(b,\Wboson)\rangle$ 
variable, both described in Section~\ref{sec:selection}. 
These have a correlation measured in data of \mbox{$\rho= -0.038$}. 
The value of $\nbtag$ in each event is determined from the 
leading six jets ordered by $\pt$.

The four regions, labelled ABCD, are identified by defining 
two bins in the number of $b$-tagged jets, $\nbtag < 2$, $\nbtag \ge 2$,
and two ranges of the $\langle\Delta\phi(b,\Wboson)\rangle$ variable, 
$\langle\Delta\phi(b,W)\rangle < 2.0$, $\langle\Delta\phi(b,\Wboson)\rangle \ge 2.0$,
as detailed in Table~\ref{tab:abcd1}.
The $\rtt$ distributions are studied for each of the defined 
regions. Region D represents the signal region (SR), and 
contains the largest fraction of $\ttbar$ events ($34.05$\%). 
Regions A, B, and C are the control regions (CR), and are 
dominated by multi-jet background events. 
Table~\ref{tab:abcd1} summarises the expected fractions 
of signal events in each of the four regions. 
Each signal fraction is estimated by comparing the total 
predicted number of signal events from $\ttbar$ simulation 
to the number of observed data events in each region.

\begin{table}[htp]
\begin{center}
\begin{tabular}{l||c|c|rcl}
\multicolumn{3}{c|}{ABCD region and definition} & \multicolumn{3}{|c}{Estimated signal fraction} \\ 
Region & $\nbtag$ & $\langle\Delta\phi(b,\Wboson)\rangle$ & \multicolumn{3}{|c}{ $\ttbar$ MC/data [\%]}\\ 
\hline 
\hline
A & $< 2$    & $\geq 2.0$ & \hspace{0.5cm} $2.06$ & $\pm$ & $0.02$ \\ 
\hline
B & $< 2$    & $< 2.0$    & $2.60$                & $\pm$ & $0.02$ \\
 \hline
C & $\geq 2$ & $\geq 2.0$ & $24.71$               & $\pm$ & $0.55$ \\ 
\hline
D & $\geq 2$ & $< 2.0$    & $34.05$               & $\pm$ & $0.57$ \\ 
\hline
\hline
\end{tabular}
\caption{Definitions and signal fractions for each of the 
four regions used to estimate the multi-jet background. 
Region D is the signal region.
The signal fraction with statistical uncertainty 
is estimated by comparing the total predicted number of 
signal events from $\ttbar$ simulation to the number of 
observed data events in each region. 
}
\label{tab:abcd1}
\end{center}
\end{table}

To obtain an unbiased estimate of the number of background 
events in each considered CR, the signal contamination is 
removed using simulated $\ttbar$ events with $\mtop = 172.5\,\textrm{GeV}$.
The method validation and the template closure described 
in Section~\ref{sec:Closure} show that the $\mtop$ dependence 
of this signal subtraction is significantly smaller than 
other uncertainties on the method, and is ignored.
The estimated background in a given bin $i$ of $\rtt$ 
for SR D ($N_{{\textrm{background}}, i}^{{\textrm{SR D}}}$) 
is given by:

\begin{equation}
N_{\text{background}, i}^{{\textrm{SR D}}} = \left(\frac{N_{\text{background}}^{{\textrm{CR C}}}}{N_{\text{background}}^{{\textrm{CR A}}}}\right){N_{\textrm{background},i}^{{\textrm{CR B}}}}.
\label{eq:SRDbackground}
\end{equation} 

The background in a given bin $i$ of the $\rtt$ spectrum 
of CR B ($N_{{\textrm{background}}, i}^{{\textrm{CR B}}}$) 
is estimated after subtraction of the signal contamination 
and after scaling by the ratio of the number events in 
control regions C (${N_{\textrm{background}}^{{\textrm{CR C}}}}$) 
and A (${N_{\textrm{background}}^{{\textrm{CR A}}}}$), 
also after signal removal.
The signal contamination present in CR C comes from improperly 
reconstructed $\ttbar$ events which form a smoothly varying 
distribution in $\rtt$.
This signal contribution in CR C is not relevant in the 
analysis, as this region only affects the normalisation of 
the distribution obtained for the multi-jet background, 
which is not used in the fit for $\mtop$ described in 
Section~\ref{sec:Top-quark measurements}.
 
Figure~\ref{fig:qcdEstimateABCD} shows the distributions 
of the masses of the $\Wboson$ boson ($m_{jj}$) and top 
quark ($m_{jjj}$) after applying the event selection, 
the $\chi^2$ approach defined in Eq.~(\ref{eq:newchi2}), 
and using the data-driven multi-jet background method.
In the figure, the reconstruction using MC events is 
said to be correct for one (or both) top quark(s) if 
each of the three jets ($j$) selected by the reconstruction 
algorithm matches to each of the three quarks ($q$) within 
a $\Delta R(j,q) < 0.3$, modulo the interchange of the 
two jets assigned to the hadronically decaying $\Wboson$ 
boson. If at least one of the jets selected by the 
algorithm is not one of the three jets matched to 
the quarks, the top quark reconstruction is classified 
as incorrect.
Finally, events where at least one quark is not matched 
uniquely to a reconstructed jet are classified as 
non-matched.
The $\rtt$ distribution obtained after using the 
data-driven multi-jet background estimation methods 
to determine the shape and normalisation 
is shown in Figure~\ref{fig:R32QCDEstimateinABCD}.
In general, good agreement between data and prediction 
is observed in all the distributions. 

\begin{figure}[htp]
  \centering
  {\resizebox{0.49\textwidth}{!}{%  
  \includegraphics{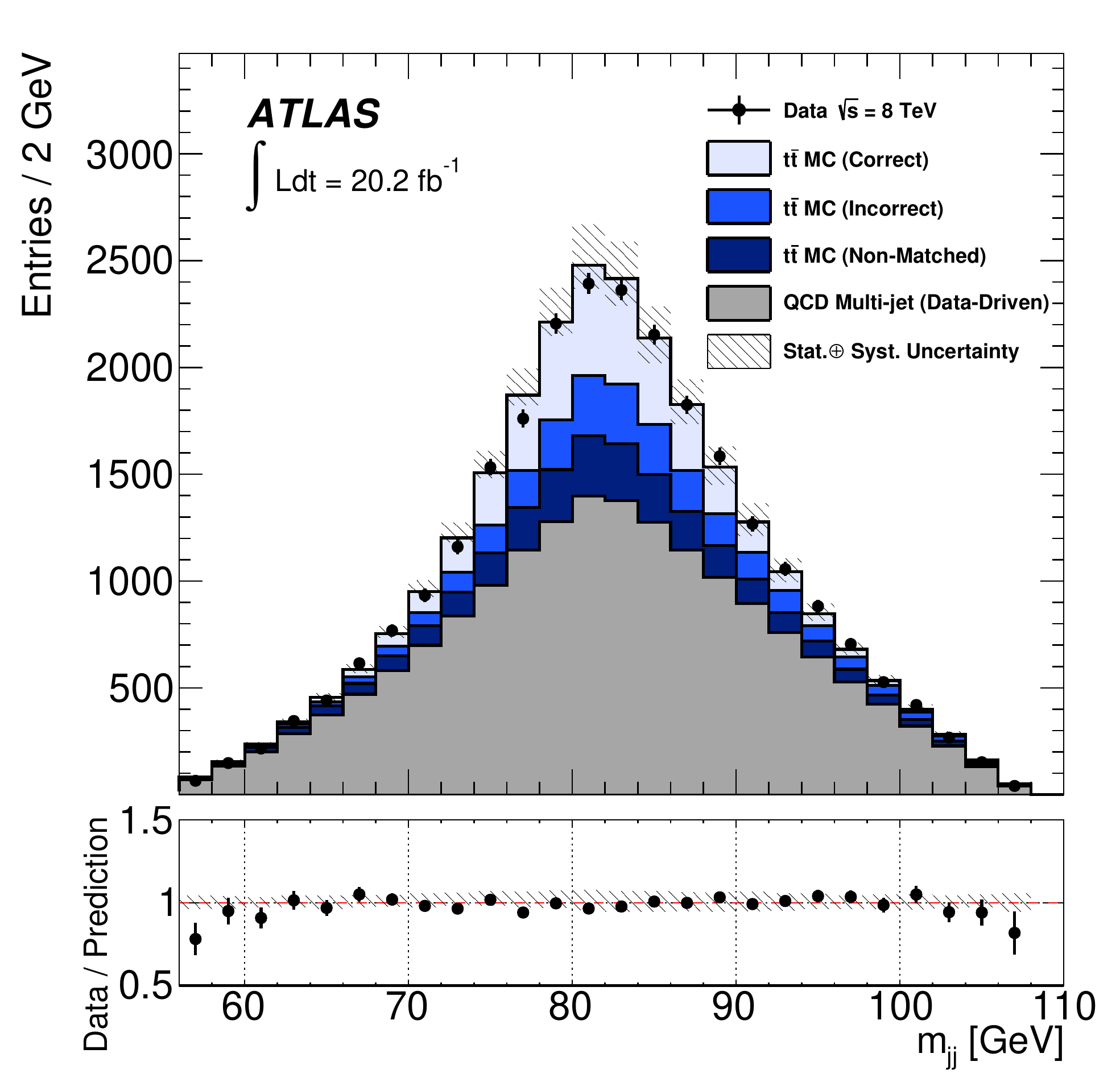}}}
  {\resizebox{0.49\textwidth}{!}{%  
  \includegraphics{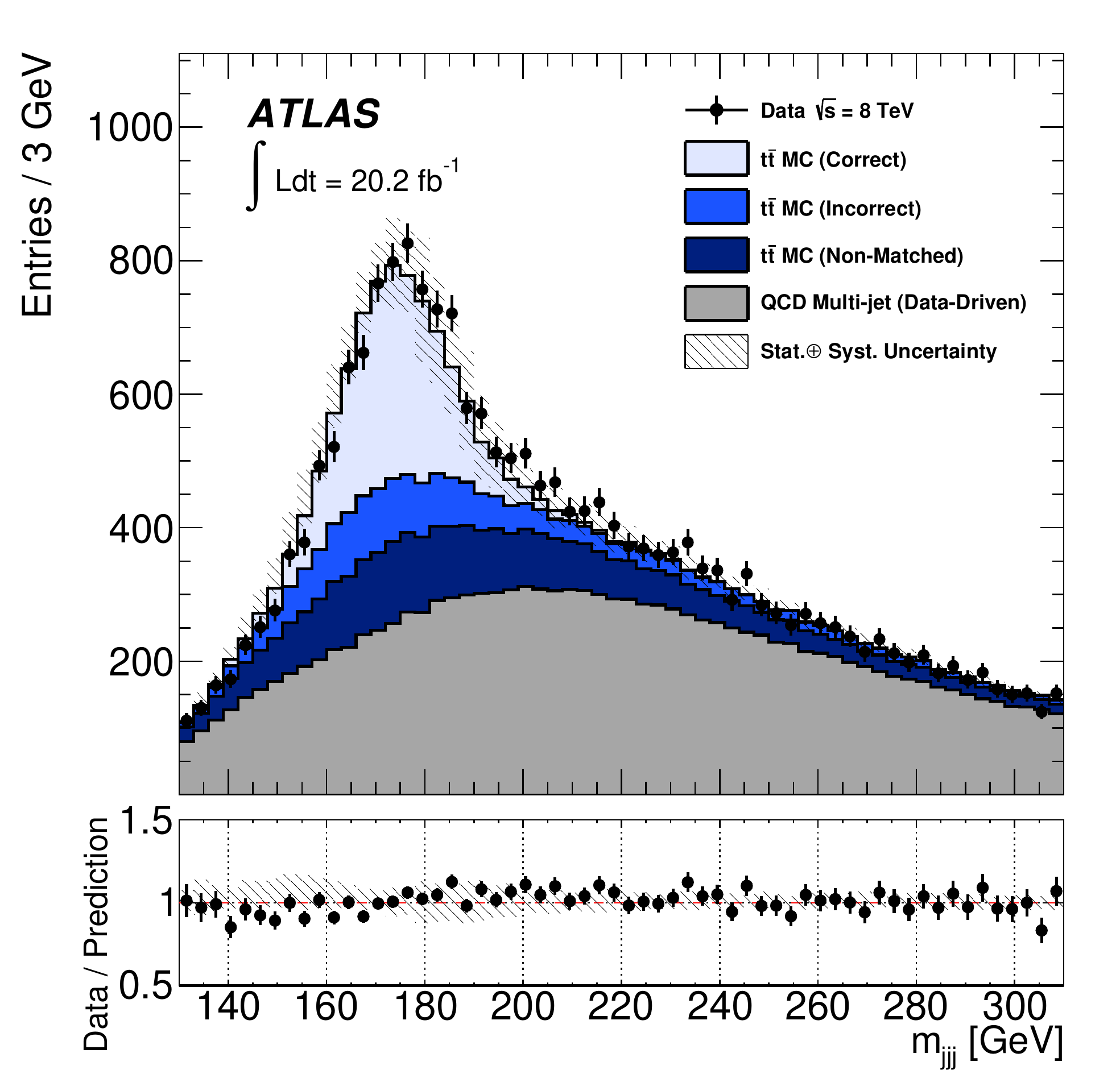}}}
  \caption{Dijet invariant mass distribution, $m_{jj}$, for $\Wboson$ 
     boson candidates (left) and three-jet invariant mass, $m_{jjj}$, 
     for top quark candidates (right) in data compared to the sum of 
     $\ttbar$ simulation and multi-jet background.
     The ratio comparing data to prediction  
     is shown below each distribution. 
     The hatched bands reflect the sum of the 
     statistical and systematic errors added in quadrature.
     The $\ttbar$ simulation corresponds to $\mtop = 172.5\,\textrm{GeV}$.}
\label{fig:qcdEstimateABCD}
\end{figure}

\begin{figure}[htp]
  \centering
  {\resizebox{0.51\textwidth}{!}{%  
  \includegraphics{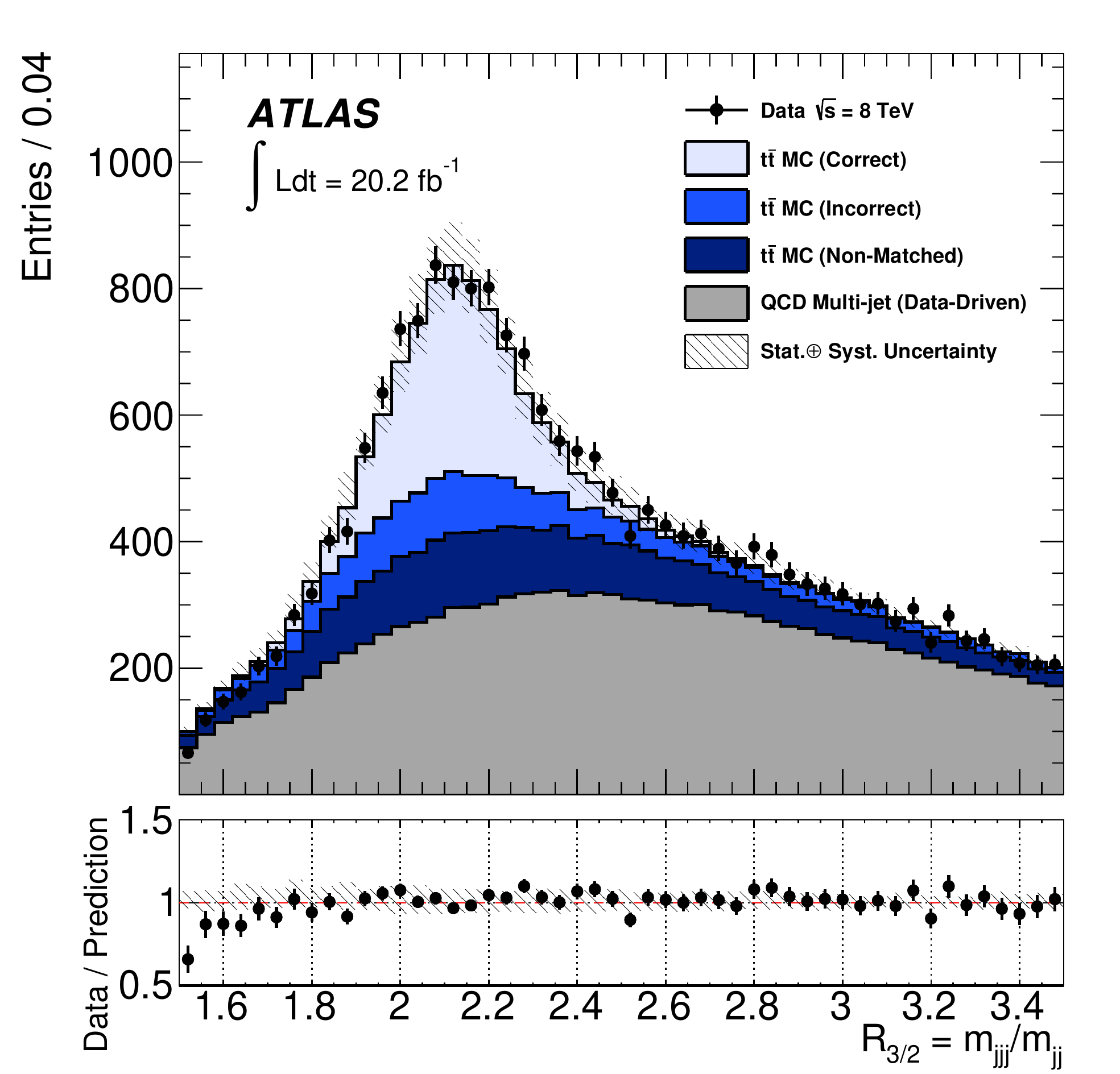}}}
  \caption{$\rtt$ distribution as obtained after applying
   the analysis event selection shown together with the 
   expected sum of $\ttbar$ simulation and 
   multi-jet background.
   The distribution is shown before the $\chi^2$ fit is 
   applied.  
   The ratio comparing data to prediction is shown below the figure.
   The hatched bands reflect the sum of the 
   statistical and systematic errors added in quadrature.
   The $\ttbar$ simulation corresponds to $\mtop = 172.5\,\textrm{GeV}$.}
  \label{fig:R32QCDEstimateinABCD}
\end{figure}

%-------------------------------------------------------------------------------
\section{Top-quark mass determination}
\label{sec:Top-quark measurements}
%-------------------------------------------------------------------------------

To extract a measurement of the top-quark mass, a 
template method with a binned minimum-$\chi^2$ approach 
is employed.  
For each $\ttbar$ event, two $\rtt$ values are obtained, 
one for each top-quark mass measurement.
To properly correct for the correlation between the two 
$\rtt$ values in each event, the statistical uncertainty 
of $\mtop$ returned from the final $\chi^2$ fit described 
later in this section is scaled up by a factor 
$\sqrt{1 + \rho} = 1.26$, where $\rho = 0.59$ 
is the correlation factor as obtained from data. 
Signal and background templates binned in $\rtt$ are 
created using the simulated $\ttbar$ events described 
in Section~\ref{sec:DataMC}, and the data-driven 
background distribution.

The top quark contribution is parameterised by a 
probability distribution function (pdf) which is the 
sum of a Novosibirsk function~\cite{Ikeda:1999aq}
and a Landau function~\cite{Koelbig:145105}. These 
describe, respectively, the signal and the combinatorial 
background.
As a first step, the $\rtt$ distributions from the 
five $\ttbar$ simulation samples with differing 
$\mtop$ are fitted separately to determine the six 
parameters for each template mass. 
The MC simulation shows that each of these parameters 
depends linearly on the input $\mtop$.
In the next step, the parameters are fitted to obtain 
the offsets and slopes of the linear $\mtop$ dependencies. 
These values are then used as inputs to a combined, 
simultaneous fit to all five $\rtt$ distributions. 
In total $12$ parameters are derived by the combined 
fit to determine the pdf. 
Figure~\ref{fig:pdf} shows the $\rtt$ distributions obtained 
using the $\ttbar$ MC samples based on the full simulation of 
the ATLAS detector and generated at three top-quark mass 
points: $167.5$, $172.5$, and $175\,\textrm{GeV}$. 
Results from the combined, simultaneous fit to all five
$\rtt$ distributions are superimposed. Shown are the functions
describing the signal and combinatorial background, respectively, 
and their sum.
The Novosibirsk mean and width parameters offer the strongest 
sensitivity to $\mtop$. 
Template distributions obtained simultaneously for three 
separate input values of $\mtop$ ($167.5$, $172.5$, and 
$177.5\,\textrm{GeV}$), highlighting the $\rtt$ shape to $\mtop$, 
are shown in Figure~\ref{fig:mtopmeasurement_mtopdep}.

\begin{figure}[htp]
  \centering
  {\resizebox{0.48\textwidth}{!}{%  
  \includegraphics{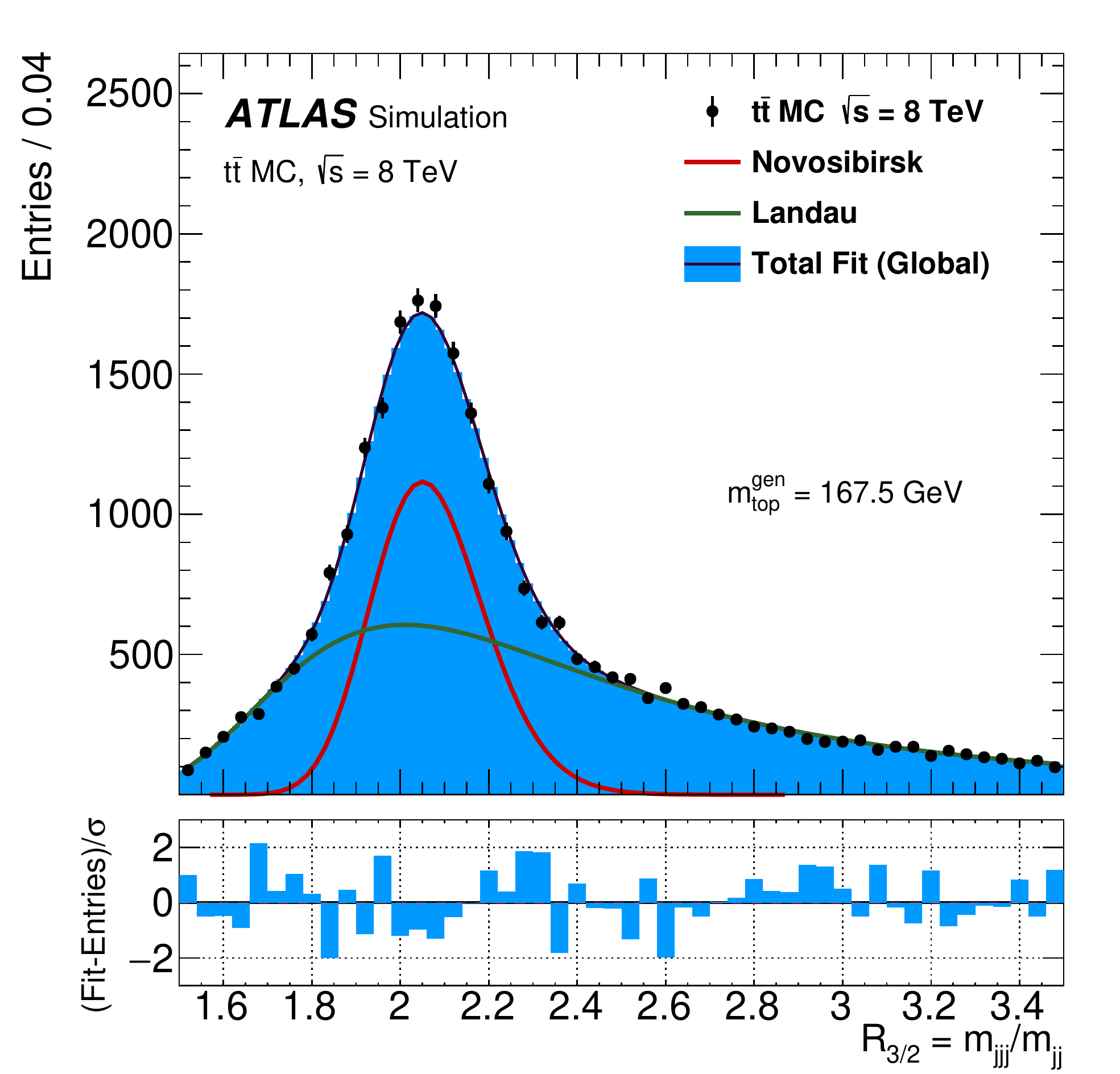}}
  \resizebox{0.48\textwidth}{!}{%  
  \includegraphics{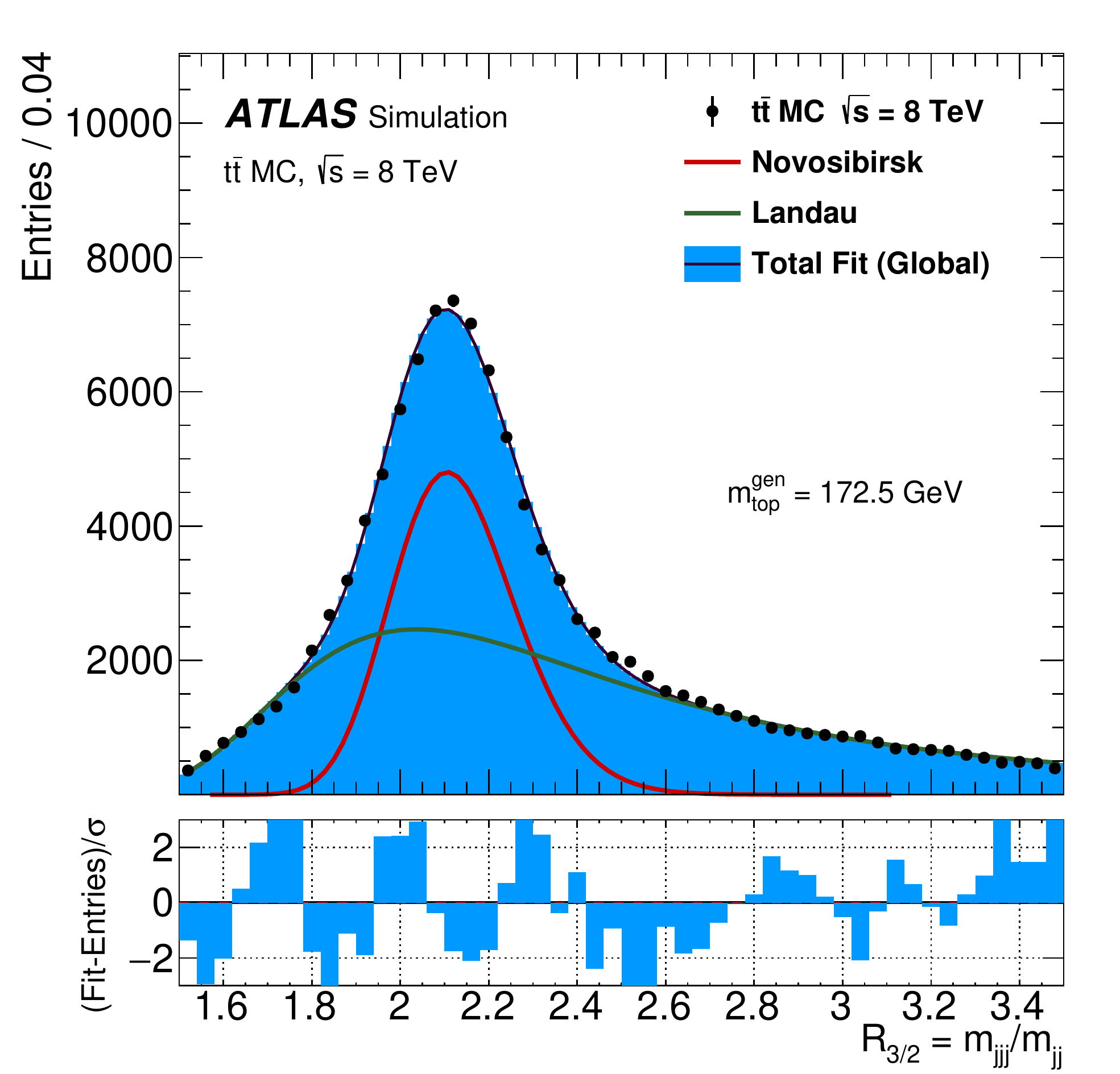}}
  \resizebox{0.48\textwidth}{!}{%  
  \includegraphics{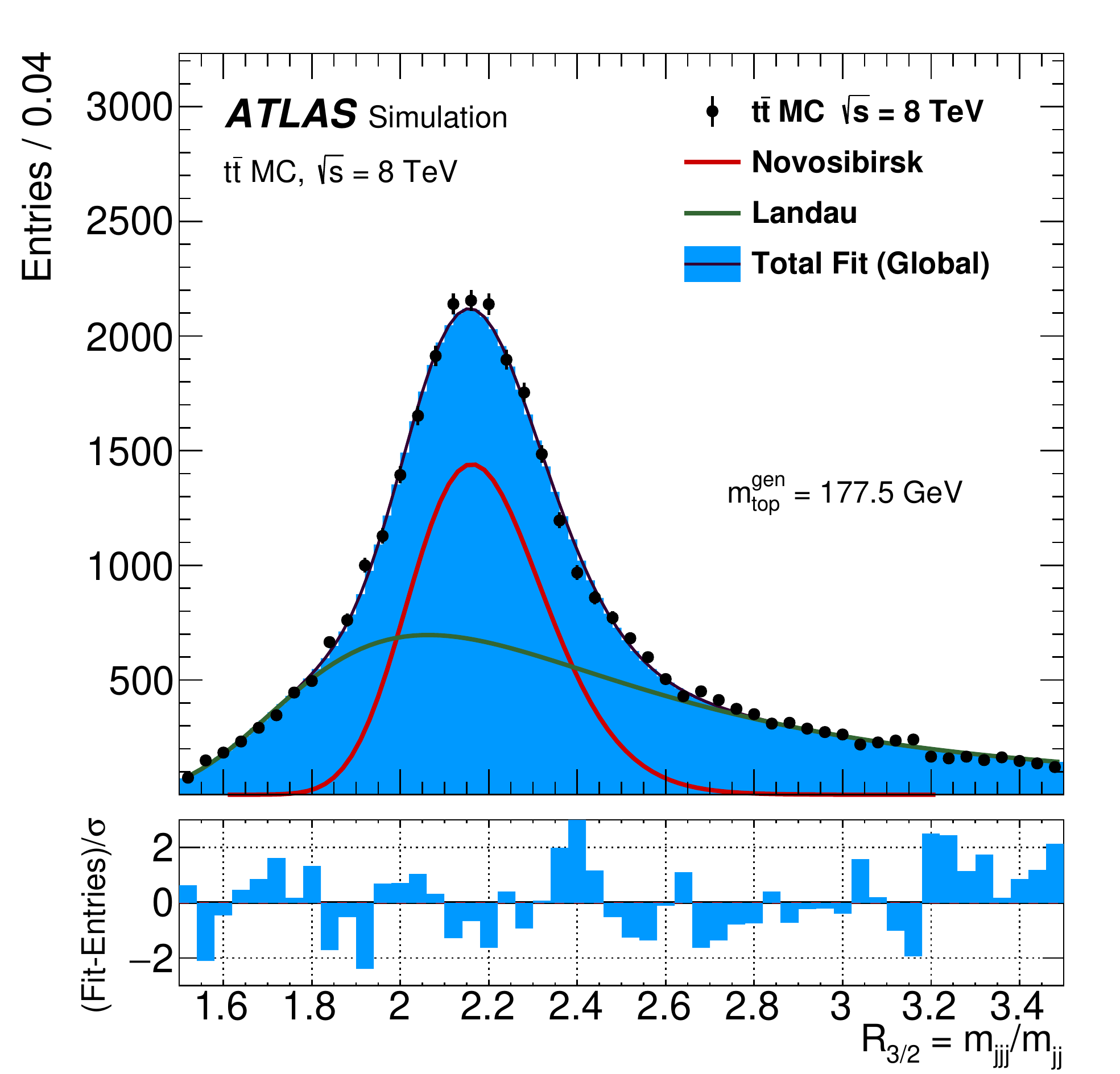}}}
  \caption{Templates for the $\rtt$ distributions for $\ttbar$ 
    MC samples generated at $\mtop$ values of $167.5$, $172.5$, and 
    $177.5\,\textrm{GeV}$, respectively.
    Results from the combined, simultaneous fit to all five
    $\rtt$ distributions are superimposed (black line with blue 
    filled area). For each distribution it consists of a 
    Novosibirsk function (red line) describing the signal part and 
    a Landau function (green dashed-line) describing the combinatorial 
    background part. Their parameters are assumed to depend linearly
    on $\mtop$. 
    The $\chi^2$ per degree of freedom obtained for 
    each of the three template distribution corresponds to $1.22$, 
    $3.98$, and $1.96$ respectively. 
    The plot under each distribution shows the residuals 
    obtained from calculating the difference between the 
    combined fit and the simulated $\rtt$ distribution normalised 
    to the statistical uncertainty for each bin individually.}
  \label{fig:pdf}
\end{figure}

\begin{figure}[htp]
  \centering
  {\resizebox{0.52\textwidth}{!}{%  
  \includegraphics{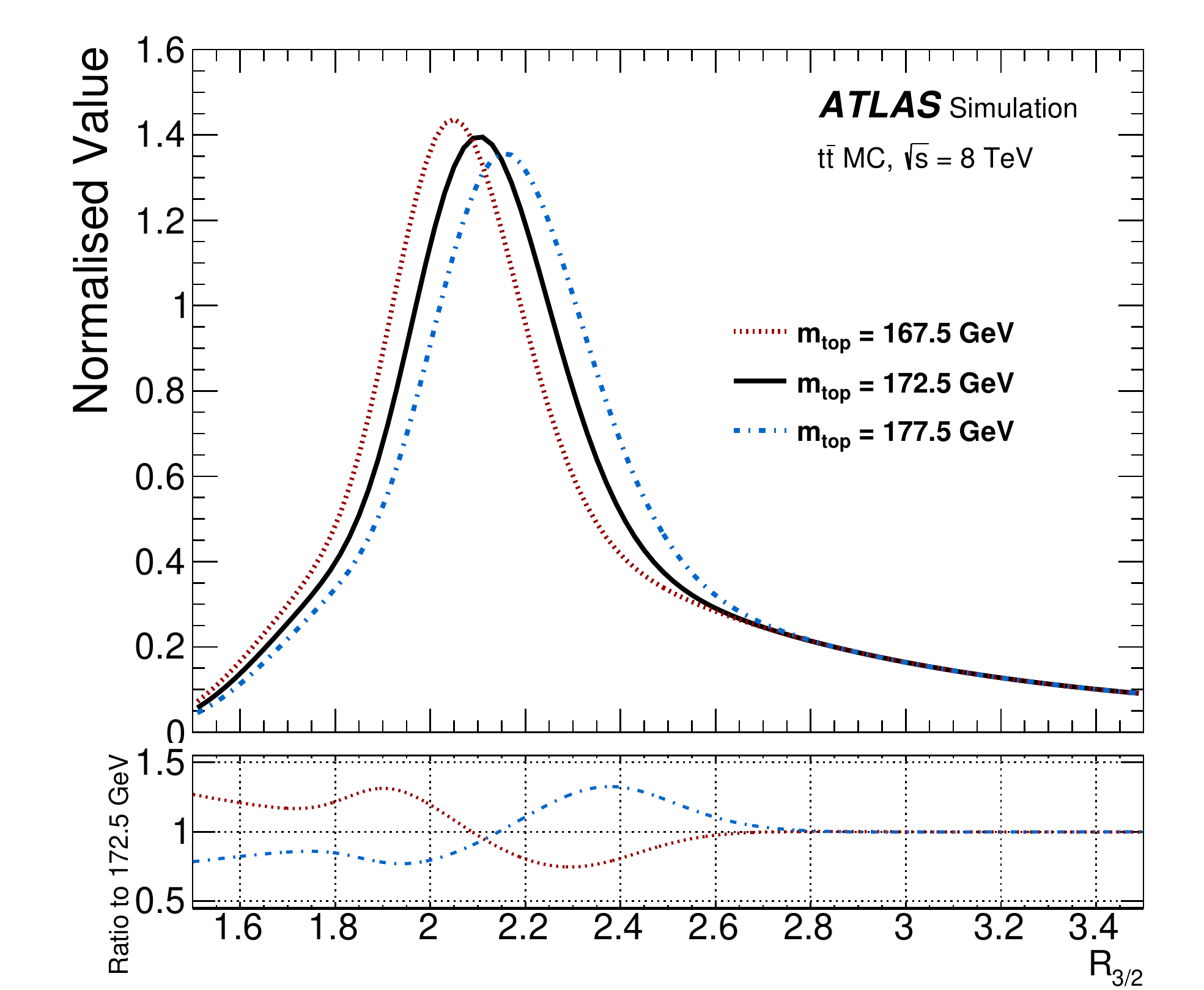}}}
  \caption{Template distributions shown simultaneously for three 
   separate input values of $\mtop$ ($167.5$, $172.5$, and 
   $177.5\,\textrm{GeV}$), 
    highlighting the sensitivity of the $\rtt$ shape to $\mtop$. 
   The plot under the distribution shows the ratio of $\mtop$ at 
   $167.5$, and $177.5\,\textrm{GeV}$ to $\mtop$ at $172.5\,\textrm{GeV}$. 
}
  \label{fig:mtopmeasurement_mtopdep}
\end{figure}

The multi-jet background template distribution obtained 
from the output of the data-driven method described in 
Section~\ref{sec:Multijet background} can be parameterised 
in a similar fashion. In this case the sum of a Gaussian 
function and a Landau function was found to be a suitable 
choice for the functional form. The background pdf requires 
five parameters.

As a final step in the parameterisation, in order to take 
properly into account the uncertainties and the correlations 
between the various signal and background shape parameters,
a more generalised version of the $\chi^2$ function is used. 
The final $\chi^2$ fit, which uses matrix algebra to
include non-diagonal covariance matrices, has the form:

\begin{equation}
\chi^2 = \sum_{i=1}^{N_{\textrm{bin}}} \sum_{k=1}^{N_{\textrm{bin}}} {\left(n_i-\mu_i\right)\left(n_k-\mu_k\right)\left[\text{\textbf{V}}_{\textrm{data}}+\text{\textbf{V}}_{\textrm{signal}}(\mtop,\fbkgd)+\text{\textbf{V}}_\textrm{bkgd}(\fbkgd)\right]^{-1}_{ik}}.
\label{EQ-MTOP-MEAUREMENT-CHI2}
\end{equation}

Here $\mtop$ and $\fbkgd$ are the two parameters which are 
left to float. The shape of the fitted multi-jet background 
parameterisation is assumed to be independent of $\mtop$ 
while the normalisation, controlled by a background fraction 
parameter, $\fbkgd$, is obtained by fitting the data distribution. 
The $\fbkgd$ is defined within the fit range of the $\rtt$ 
distribution: $1.5 \le \rtt < 3.5$.                                                                      
The term $n_i$ in Eq.(\ref{EQ-MTOP-MEAUREMENT-CHI2}) 
corresponds to the number of entries in bin $i$ in the 
$\rtt$ data distribution, whereas $\mu_i$ corresponds 
to the estimated total number of signal and background 
entries. The term $\textbf{V}_{\text{data}}$ is the 
$N_{\textrm{bin}} \times N_{\textrm{bin}}$ diagonal 
data covariance matrix with $V_{ik} = \delta_{ik} n_i$, 
which accounts for the statistical uncertainty in each 
bin $i$. Similarly, $\textbf{V}_\textrm{signal}$ and 
$\textbf{V}_\textrm{bkgd}$ are 
$N_{\textrm{bin}} \times N_{\textrm{bin}}$ non-diagonal 
covariance matrices which account for the signal and 
background shape parameterisation uncertainties and 
their correlations.
In the $\rtt$ distribution which has a total number of 
data entries $N_d$, and a given bin width $w_{\textrm{bin}}$, 
the number of estimated entries in bin $i$, $\mu_i$, is 
given by:

\begin{equation}
\mu_i \left( \mtop,\fbkgd \right) = w_{\textrm{bin}} N_d \left[ \left(1-\fbkgd \right) P_{\textrm{S}} \left(R_{\textrm{3/2},i} | \mtop \right)+\fbkgd P_{\textrm{B}} \left(R_{\textrm{3/2},i} \right) \right]
\label{EQ-MUI}
\end{equation}

where $P_{\textrm{S}}$ and $P_{\textrm{B}}$ are the probability 
density functions for the signal and background, respectively.

%-------------------------------------------------------------------------------
\section{Method validation and template closure}
\label{sec:Closure}
%-------------------------------------------------------------------------------
%
To validate the method employed to extract $\mtop$ 
from the $\rtt$ data distribution and to check for any 
potential bias, a series of pseudo experiments are 
performed. For each of the five simulated $\mtop$ 
samples a total of $2500$ pseudo experiments generating 
a distribution of the $\rtt$ variable are 
produced~\footnote{This value of $2500$ is also used when 
performing pseudo experiments to estimate the systematic 
uncertainties.}.  
Two scenarios are investigated: in the first one, 
events are drawn randomly from template $\rtt$ 
distributions; in the second scenario, events are drawn 
directly from the signal and background shapes.
In each scenario the nominal values of all signal 
and background shape parameters are used, and only 
two parameter values, $\mtop$ and $\fbkgd$, are returned 
from the minimisation procedure. 
For all five top-quark mass MC samples, the same 
multi-jet background distribution is used for drawing 
pseudo events.

The value of $\mtop$ obtained from each pseudo experiment 
($\mtop^{\textrm{meas}}$) is used to fill a distribution 
of the difference between these values and 
the values $\mtop^{\textrm{gen}}$ used for event generation.
This distribution is then fitted with a Gaussian function,
giving estimators for the Gaussian mean and width 
parameters, each with their respective uncertainties.
The uncertainty in the fitted mean is corrected for the 
oversampling that is induced by drawing from template 
distributions produced using a finite number of MC 
events~\cite{Barlow1}.
The fitted mean $\left<\left(\mtop^{\textrm{meas}}-\mtop^{\textrm{gen}}\right)\right>$, 
referred to as the ``difference mean'', is shown in Figure~\ref{fig:diff_mtop}.
 Fitting the difference mean for the five top-quark mass 
samples with a linear function gives an $\mtop^{\textrm{gen}}$-independent 
bias of $0.08 \pm 0.06\,{\textrm{GeV}}$.
The treatment of this small bias is further discussed in 
Section~\ref{sec:systematics_nonclosure}.

Pull distributions are constructed in an analogous way, 
where the pull in each pseudo experiment is defined as:

\begin{equation}
{\text {Pull}} = \left(\mtop^{\textrm{meas}}- \mtop^{\textrm{gen}}\right)/\delta \mtop,  
\label{eq:Corr_PullMean}
\end{equation}
where $\delta \mtop$ is the statistical uncertainty of the 
$\mtop$ parameter obtained from the fit of the pseudo experiment. 
The correction that takes into account the correlation between 
two $\rtt$ values in each event, described in 
Section~\ref{sec:Top-quark measurements}, is not applied here, 
as the values of $\rtt$ drawn for pseudo experiments are 
uncorrelated.
The pull distribution for an unbiased measurement has a 
mean of zero and a standard deviation of unity.
A fitted pull mean value of $0.19 \pm 0.13$ and a 
fitted pull width of $0.98 \pm 0.01$ are obtained, 
which shows that the uncertainty determination is 
unbiased.

\begin{figure}[htp]
\centering
    \includegraphics[scale=0.45]{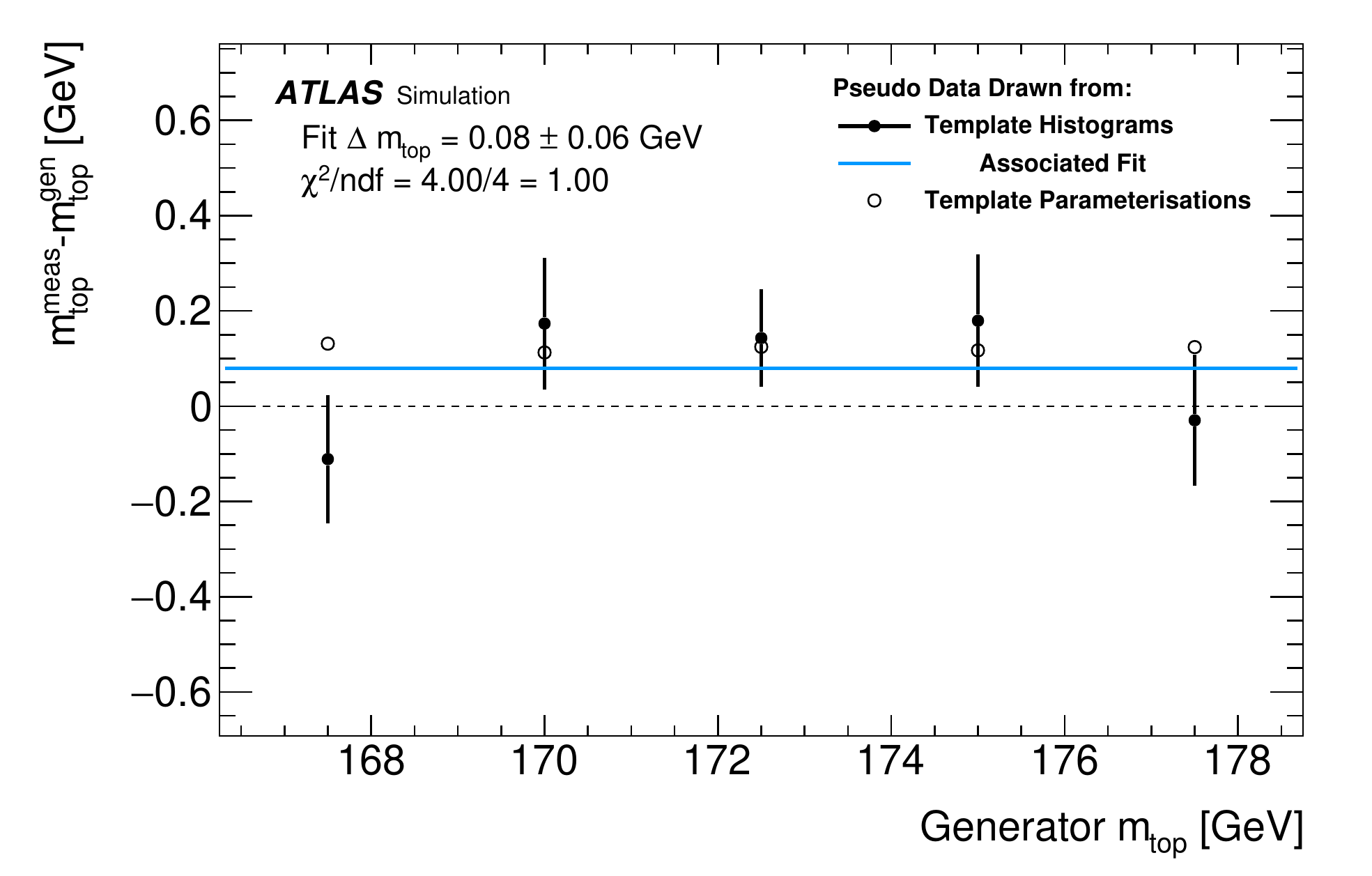}
\caption{The difference mean, $\left(\mtop^{\textrm{meas}}-\mtop^{\textrm{gen}}\right)$, 
 based on the results of a fit to a single Gaussian function. The black markers 
 correspond to cases where the pseudo events were drawn from the 
 $\rtt$ histograms, and the open marker points where pseudo events 
 were drawn from the parameterisations.  
 The solid blue line corresponds to a polynomial fit 
 to the five black markers and their corrected 
 uncertainties.}
\label{fig:diff_mtop} % caption for the whole figure
\end{figure}

%-------------------------------------------------------------------------------
\section{Systematic uncertainties}
\label{sec:systematic uncertainties}
%-------------------------------------------------------------------------------

This section outlines the various sources of systematic 
uncertainty in $\mtop$ which are summarised in 
Table~\ref{tab:sys_all}.
All sources are treated as uncorrelated. Individual 
contributions are symmetrised and the total uncertainty is 
taken as the sum in quadrature of all contributions.

The majority of the systematic uncertainties are assessed 
by varying the $\ttbar$ MC sample to reflect the uncertainty 
from each of these sources.
Pseudo experiments are constructed from the varied sample, 
which are then passed through the analysis chain; the change 
in the result relative to that obtained from the nominal MC 
sample is evaluated.
Exceptions to this are described in the following subsections. 
To facilitate a combination with other results, each systematic 
uncertainty is assigned a statistical uncertainty, taking into 
account the statistical correlation of the considered samples. 
Following Ref.~\cite{Barlow:2002yb}, the systematic uncertainties 
listed in Table~\ref{tab:sys_all} are calculated 
independently of the statistical uncertainties of the values.

In what follows, each source of systematic uncertainty is briefly 
described. 
These are broken down into three categories. The first category, 
theory and modelling uncertainties, is associated with the 
simulation of the signal events.  
The second set of uncertainties is related to the analysis method.
These involve uncertaintie due to the way that the analysis was 
performed, including the choice of a template method, the 
background modelling, and the final $\mtop$ extraction procedure. 
Finally a third category, calibration- and detector-related 
uncertainties involves uncertainties coming from the standard 
calibrations of physics objects.

\begin{table}[htp]
\begin{center}
\begin{tabular}{l||c}
\itshape{Source of uncertainty}  &  $\Delta \mtop$ [${\textrm{GeV}}$]  \\ 
\hline 
\hline 
Monte Carlo generator        &  $0.18 \pm 0.21$ \\ 
Hadronisation modelling      &  $0.64 \pm 0.15$ \\ 
Parton distribution functions&  $0.04 \pm 0.00$ \\ 
Initial/final-state radiation&  $0.10 \pm 0.28$ \\ 
Underlying event             &  $0.13 \pm 0.16$ \\ 
Colour reconnection          &  $0.12 \pm 0.16$ \\ 
\hline
Bias in template method      &  $0.06$          \\ 
Signal and bkgd parameterisation& $0.09$        \\ 
Non all-hadronic $\ttbar$ contribution& $0.06$  \\ 
ABCD method {\itshape{ vs.}} ABCDEF method & $0.16$ \\ 
\hline 
Trigger efficiency           &  $0.08 \pm 0.01$ \\ 
Lepton/$\met{}$ calibration  &  $0.02 \pm 0.01$ \\ 
Overall flavour-tagging      &  $0.10 \pm 0.00$ \\ 
Jet energy scale (JES)       &  $0.60 \pm 0.05$ \\ 
b-jet energy scale (bJES)    &  $0.34 \pm 0.02$ \\ 
Jet energy resolution        &  $0.10 \pm 0.04$ \\ 
Jet vertex fraction          &  $0.03 \pm 0.01$ \\ 
\hline 
\hline 
\textbf{Total systematic uncertainty} &  $1.01$ \\
\textbf{Total statistical uncertainty}&  $0.55$ \\
\hline 
\hline 
\textbf{Total uncertainty}  & $1.15$ \\
\hline
\hline 
\end{tabular}
\caption[Summary of sources of statistical and systematic 
uncertainties in $\mtop$ and $\fbkgd$]{Summary of all 
sources of statistical and systematic uncertainties in 
the measured values of the top-quark mass.  
Totals are evaluated by means of a sum in quadrature 
and assuming that all contributions are uncorrelated. 
The uncertainties are subdivided into three categories: 
theory and modelling uncertainties, method-related 
uncertainties, and calibration- or detector-related 
uncertainties, as described in the text. 
Adjacent to each of the quoted systematic variations 
in $\mtop$ is its associated statistical uncertainty. 
The ABCDEF method is further described in 
Section~\ref{sec:background_number_of_regions}
and in Ref.~\cite{Aad:2014zea}.
The quoted statistical uncertainty is corrected for 
the correlation between the two $\rtt$ measurements 
of each event.
}
\label{tab:sys_all}
\end{center}
\end{table}

%-------------------------------------------------------------------------------
\subsection{Theory and modelling uncertainties}
\label{sec:systematics_theory}
%-------------------------------------------------------------------------------

\textbf{Monte Carlo generator:}
\label{sec:systematics_mcgen}
In order to assess the impact on the $\mtop$ measurement 
due to the choice of MC generator, the results of pseudo 
experiments using two different AFII simulated samples 
are compared: one sample produced using \textsc{POWHEG-BOX} 
as the MC generator and a second sample using 
\textsc{MC@NLO}~\cite{Frixione:2002ik}. 
Both samples use \textsc{Herwig 6.520.2}~\cite{Corcella:2000bw} 
with the \textsc{AUET2} tune to model the parton shower, 
hadronisation and underlying event, in contrast with the 
nominal signal MC where \textsc{Pythia  6.427} is used.  
The absolute difference of $0.18\,\textrm{GeV}$ between the 
resulting average $\mtop$ parameter returned from the fits
is accounted for as the uncertainty.

\textbf{Hadronisation modelling:}
\label{sec:systematics_had}
To quantify the expected change in the measured $\mtop$ 
value due to a different choice of hadronisation model,
pseudo experiments are performed for two independent MC samples 
both employing \textsc{POWHEG-BOX} AFII simulation to generate 
the all-hadronic $\ttbar$ events but differing in their choice 
of hadronisation model. 
In the first case, \textsc{Pythia  6.427}~\cite{Sjostrand:2006za} 
is used to model the parton shower, hadronisation and underlying 
event with the \textsc{Perugia 2012} tunes~\cite{Skands:2010ak}, 
while in the second case, \textsc{Herwig 6.520.2} with 
the \textsc{AUET2} tune~\cite{Corcella:2000bw} is used. 
The absolute difference of $0.64\,\textrm{GeV}$ between the 
average $\mtop$ values obtained in the two cases is accounted 
for as the systematic uncertainty.

\textbf{Parton distribution functions:}
\label{sec:systematics_pdf}
A variety of PDF sets are investigated in order to assess 
the impact of the choice of \textsc{CTEQ10}~\cite{Lai:2010vv,Gao:2013xoa}, 
the default PDF set used in the nominal measurement. 
There are a total of $53$ distinct sets for the \textsc{CTEQ} 
PDFs. In addition there are $101$ distinct \textsc{NNPDF23}~\cite{Ball:2012cx} 
PDF sets and $41$ distinct \textsc{MSTW2008}~\cite{Martin:2009iq,Martin:2009bu} 
PDF sets to consider, giving a total of $195$ distinct sets to compare. 
Simulated \textsc{POWHEG-BOX}+\textsc{Herwig}~\cite{Nason:2004rx,Frixione:2007vw,Alioli:2010xd,Corcella:2000bw} 
events are used for the comparison. 
The individual PDF uncertainty contributions are evaluated 
according to set-dependent procedures as described in 
Ref.~\cite{Botje:2011sn} for \textsc{CT10}~\cite{Gao:2013xoa,Lai:2010vv}, 
for \textsc{MSTW}~\cite{Ball:2012cx}, and 
for \textsc{NNPDF}~\cite{Martin:2009iq,Martin:2009bu}.  
To determine the final systematic uncertainty, the quantity 
$\mtop$ is calculated for each of the three sets, where 
$\mtop$ is the measured value from the central reference 
sample of the corresponding PDF set.
Half of the difference between the largest and the smallest 
of these $\mtop$ values is quoted as the symmetrised final 
uncertainty, and is $0.04\,\textrm{GeV}$.

\textbf{Initial-state and final-state radiation:}
\label{sec:systematics_ifsr}
Varying the amount of initial- and final-state radiation 
(ISR and FSR) can have an impact on the number of 
reconstructed jets, which in turn can affect the overall 
measurement of the top-quark mass. In order to quantify 
the sensitivity of the measurement to ISR/FSR, two alternative 
\textsc{POWHEG-BOX} plus \textsc{Pythia 6.427}~\cite{Sjostrand:2006za} 
AFII samples are used. The first sample has the $h_{\textrm{damp}}$ 
parameter~\cite{ATL-PHYS-PUB-2014-005} set to 2$\mtop$, 
the factorisation and renormalisation scale~\footnote{The default
\textsc{POWHEG-BOX} factorisation and renormalisation scales are 
set to $\sqrt{\mtop^2 + \pt^2}$} decreased by a factor of $0.5$ 
and uses the \textsc{Perugia 2012 radHi} tune~\cite{Skands:2010ak}, 
giving more parton shower radiation. 
The second sample has the \textsc{Perugia 2012 radLo} tune, 
$h_{\textrm{damp}} = \mtop$ and the factorisation and renormalisation 
scale increased by a factor of $2$, giving less parton shower radiation. 
Half of the absolute difference between the measured $\mtop$ values 
from the pseudo experiments is quoted as the corresponding systematic 
uncertainty and is $0.10\,\textrm{GeV}$.

\textbf{Underlying event:}
\label{sec:systematics_ue}
Additional semi-hard multiple parton interactions (MPI) 
present in the hard-scattering can change the kinematics 
of the underlying event.
The number of such additional semi-hard MPI is a 
\textsc{Perugia 2012} tunable parameter~\cite{Skands:2010ak} 
in the \textsc{Pythia 6.427} generator~\cite{Sjostrand:2006za}.
Simulated $\ttbar$ AFII events were produced  with an increased 
number of semi-hard MPI (\textsc{Perugia 2012 mpiHi}) in order to 
assess the potential impact on the final measurement. 
The absolute difference between the results of these pseudo 
experiments and the one using the nominal simulated AFII sample 
is quoted as the systematic uncertainty and is $0.13\,\textrm{GeV}$.

\textbf{Colour reconnection:}
\label{sec:systematics_cr}
When simulating AFII signal events using \textsc{Pythia 6.427}~\cite{Sjostrand:2006za} 
for the parton shower and hadronisation modelling, there is a 
tunable parameter associated with the colour reconnection strength 
due to the colour flow along parton lines in the strong-interaction 
hard-scattering process. 
An alternative AFII $\ttbar$ sample uses the \textsc{Pythia Perugia 2012 loCR} 
tune~\cite{Skands:2010ak}, which corresponds to reduced colour 
reconnection strength. The absolute difference of $0.12\,\textrm{GeV}$ 
between the results of these pseudo experiments and the average 
$\mtop$ value obtained using the nominal \textsc{Pythia 6.427} 
$\ttbar$ events is quoted as the systematic uncertainty.

%-------------------------------------------------------------------------------
\subsection{Method-dependent uncertainties}
\label{sec:systematics_method}
%-------------------------------------------------------------------------------

\textbf{Bias in template method:}
\label{sec:systematics_nonclosure}
Based on the results of the closure tests, a small bias 
is observed in the extracted top-quark mass. 
By drawing pseudo events from the parameterisations an 
offset of about $80\,\textrm{MeV}$ in the mass difference 
($\mtop^{\textrm{meas}} - \mtop^{\textrm{gen}}$) 
is present (see Figure~\ref{fig:diff_mtop}).
The offset does not exhibit a dependence on the 
generator's $\mtop$ value. 
For this reason the parameter value returned 
from a fit to the average bias from pseudo experiments 
across $\mtop^{\textrm{gen}}$ is subtracted from the 
final $\mtop$ value as measured in data. 
The final value of $\mtop$ quoted in this analysis includes 
this subtraction. The uncertainty in this fitted offset is 
then quoted as the systematic uncertainty of $0.06\,\textrm{GeV}$ 
associated with the template method's non-closure.

\textbf{Signal and background parameterisation:}
\label{sec:systematics_sbparam}

To extract $\mtop$ as described in Section~\ref{sec:Top-quark measurements},
the uncertainties in the shape parameters of the $\rtt$ 
observable for the signal contributions are included in the 
$N_{\textrm{bin}} \times N_{\textrm{bin}}$ covariance matrices 
which enter into the $\chi^2$ minimisation used to extract 
$\mtop$ (see Eq.~(\ref{EQ-MTOP-MEAUREMENT-CHI2})).
Omitting these contributions would yield a simplified definition 
of the $\chi^2$ variable:

\begin{equation}
\chi^2 = \sum_{\textrm{bin}\ i}^{N_{\textrm{bin}}} \sum_{\textrm{bin}\ k}^{N_{\textrm{bin}}} {\left(n_i-\mu_i\right)\left(n_k-\mu_k\right)\left[\text{\textbf{V}}_\textrm{data}\right]^{-1}_{ik}} =  \sum_{\textrm{bin}\ i}^{N_{\textrm{bin}}} \frac{{\left(n_i-\mu_i\right)^2}}{n_i}
\label{EQ-11-CHI2-SBPARAM}
\end{equation}

which can be recognised as the standard definition of the 
$\chi^2$ variable for a least-squares fit assuming only a 
diagonal covariance matrix.  
The fit to the data distribution is repeated using this 
simplified definition of the $\chi^2$ variable. 
This results in a slightly modified returned value of the 
$\mtop$ parameter and a smaller statistical uncertainty. 
The difference in quadrature of $0.09\,\textrm{GeV}$ between the 
final statistical uncertainty returned from the original 
minimisation and this modified value is quoted as the 
uncertainty in the signal and background parameterisation.

\textbf{Inclusion of non-all-hadronic $\ttbar$ background:}
\label{sec:systematics_nonallhad}
A number of event selection requirements, such as the lepton 
veto and the requirement that $\met{} < 60\,\textrm{GeV}$, result 
in a large suppression of background contributions arising 
from non-all-hadronic $\ttbar$ events. 
The estimated fractional contribution from such events in 
the final signal region is below $3$\%, and is not considered 
in the nominal case. 
Pseudo experiments are performed by drawing events from the 
nominal signal distribution but from a modified background, 
now consisting of QCD events and $\ttbar$ events with at 
least one leptonic $\Wboson$ boson decay. 
The absolute difference of $0.06\,\textrm{GeV}$ between the 
average $\mtop$ value obtained in this way and that from 
the nominal case is quoted as a systematic uncertainty.

\textbf{Variation in the number of control regions:}
\label{sec:background_number_of_regions}

A variation of the background estimation procedure is 
considered in which six distinct regions, rather than 
four, are defined to estimate the multi-jet background. 
This is done by allowing three different values of 
$\nbtag$: $0$, $1$, or $\ge 2$. 
Events can then be separated into the six differing 
regions as in the nominal analysis. 
As in the nominal case the number of $b$-tagged jets 
in an event considers only the leading six jets, ordered 
by $\pt$. The values of the second ABCD variable, 
$\langle \Delta\phi(b,\Wboson)\rangle$, are unchanged 
from the nominal case. One reason for considering 
this alternative is that the inclusion of a larger 
number of control regions could potentially provide 
sensitivity to different physics processes. 
Additionally, the systematic uncertainty contribution 
arising from uncertainties in the $b$-tagging scale 
factors could differ between these methods.

With a total of six regions shown in Table~\ref{tab:abcdef} 
the background estimation technique remains similar to 
that using four regions.

\begin{table}[htp]
\begin{center}
\begin{tabular}{c||c|c|c|c|c|c}
Region     &A & B &  C & D & E & F   \\ 
\hline 
\hline
$\nbtag$ & $0$          &  $0$         &  $1$       &  $1$       &  $\ge 2$  & $\ge 2$ \\   
$\langle\Delta\phi(b,\Wboson)\rangle$ & $\ge 2$ & $< 2$ & $\ge 2$ & $< 2$ & $\ge 2$ & $< 2$ \\ 
 \hline
\hline
\end{tabular}
\caption{Definitions for each of the six regions ABCDEF used to estimate the multi-jet background.}
\label{tab:abcdef}
\end{center}
\end{table}

The final SR is labelled F.  The new region D, together with region B, is now 
used to predict the shape of the multi-jet background in SR F, whereas CR A, C, 
and E set the multi-jet background normalisation~\cite{Aad:2014zea}. 
Pseudo experiments are performed by drawing background events from the 
modified multi-jet distribution in the final signal region. 
The absolute difference of $0.16\,\textrm{GeV}$ between this and the nominal 
case is quoted as the systematic uncertainty.

%-------------------------------------------------------------------------------
\subsection{Calibration- and detector-related uncertainties}
\label{sec:systematics_detector}
%-------------------------------------------------------------------------------

\textbf{Trigger efficiency:}
\label{sec:systematics_trig}

The trigger efficiency obtained using simulated signal 
events~\cite{Nason:2004rx,Frixione:2007vw,Alioli:2010xd,Sjostrand:2006za,Skands:2010ak} 
is compared to an equivalent distribution obtained using data,
which results in a small observed discrepancy.
The data here are expected to consist primarily of 
multi-jet events. It is expected that some true kinematic 
differences give rise to the difference observed between 
the data and MC trigger efficiencies. 
In order to obtain a conservative uncertainty, it is assumed 
that the difference represents a mis-modelling of the data 
by the trigger simulation. The simulated events are assigned 
a $\pt$-dependent trigger efficiency correction such that 
the corrected MC and data trigger efficiencies agree.
Pseudo experiments are performed by drawing signal events 
from the modified $\rtt$ distribution with the trigger SFs 
applied, and the $0.08\,\textrm{GeV}$ absolute difference from 
the nominal case is quoted as a conservative uncertainty 
on $\mtop$ due to the trigger efficiency.

\textbf{Pile-up reweighting scale:}
\label{sec:systematics_musc}

The distribution of the average number of interactions 
per bunch crossing, denoted by $\langle\mu\rangle$, is 
known to differ between data and simulation. 
Simulated events are reweighted so that $\langle\mu\rangle$ 
matches the value observed in data. In order to assess 
the impact on the final result, pseudo experiments are 
performed in which the reweighting scale is shifted up 
and down according to its uncertainty, and the fit 
procedure is repeated. A negligible maximum change of 
$0.01\,\textrm{GeV}$ in $\mtop$ is found as the symmetrised 
up/down uncertainty.

\textbf{Lepton and $\met{}$ soft-term calibrations:}
\label{sec:systematics_lepmet}

Uncertainties in the calibration scales and in the 
resolutions of the lepton ($e/\mu$) four-vector 
objects~\cite{Aad:2014fxa,Aad:2014rra,Aad:2014nim}
can potentially lead to small differences in the 
event selection or the jet--quark assignment in the 
top reconstruction algorithm. Similarly, 
small uncertainties in $\mtop$ can be expected due 
to the uncertainties in the scale and resolution 
of the $\met{}$ soft term~\cite{Aad:2016nrq,ATLAS-CONF-2014-019}. 
The $\met$~soft term is varied according to these 
uncertainties and pseudo experiments are performed 
with the modified MC events. In the case of the 
muon-related uncertainties, Gaussian smearing is 
performed to assess the impact  on the final result. 
The maximum absolute deviation from the reference 
$\mtop$ value is taken as the uncertainty in each 
case, and these are added in quadrature to obtain 
a single value of $0.02\,\textrm{GeV}$ for all 
lepton- and $\met{}$-related scale and resolution 
uncertainties.

\textbf{Flavour-tagging efficiencies:}
\label{sec:systematics_btag}
In the validation of the flavour-tagging algorithms, 
the differences between tagging efficiencies and mis-tag 
rates evaluated in data and simulation are removed
by applying scale factor (SF) weights to the simulated 
events.
The uncertainties in the flavour-tagging SFs are calculated 
separately for the $b$-tagging SFs, the $c/ \tau$-tagging 
SFs, and the overall mis-tag SFs~\cite{ATLAS-CONF-2014-046}. 
The uncertainties in the flavour-tagging SFs are split 
into various components. 
The full covariance matrix between the various bins of 
jet transverse momentum is built and decomposed into 
eigenvectors. 
Each eigenvector corresponds to an independent source 
of uncertainty, each with an upward and a downward 
fluctuation, and the resulting total systematic 
uncertainty is $0.10\,\textrm{GeV}$.

\textbf{Jet energy scale:}
\label{sec:systematics_jes}
The different contributions to the total JES uncertainty 
are estimated individually as described in Ref.~\cite{Aad:2014bia}.
For each component the resulting differences from the up 
and down variations, corresponding to one-standard-deviation  
relative to the nominal JES, are quoted separately. 
The total uncertainty for each contribution is taken as 
half of the absolute difference between the up and down 
variation. In case both the up and down variations result 
in a change in the parameter in the same direction, the 
largest absolute difference (either from the up or down 
variation) is taken as the symmetrised uncertainty. 
The total JES uncertainty is the sum in quadrature of 
all subcontributions, and is $0.60\,\textrm{GeV}$.  
This includes all but the $b$-jet energy scale 
contribution, which is quoted separately and discussed 
below.

\textbf{$b$-jet energy scale:}
\label{sec:systematics_bjes}
The reconstructed top quark four-momenta are sensitive to 
the energy scale of jets initiated by $b$-quarks, 
particularly as a result of choices in the fragmentation 
modelling. Based on the uncertainties associated with the 
$b$-jet energy scale~\cite{ATLAS-CONF-2013-002}, a similar 
up/down variation procedure is performed using pseudo 
experiments and the quoted systematic uncertainty of 
$0.34\,\textrm{GeV}$ is half the absolute difference between 
the two variations.

\textbf{Jet energy resolution:}
\label{sec:systematics_jer}
An eigenvector decomposition strategy similar to that followed 
for the JES and the flavour-tagging systematic uncertainties is 
used for the determination of jet energy resolution (JER) systematic 
uncertainties~\cite{ATLAS-CONF-2015-037}.
The final quoted JER systematic uncertainty is $0.10\,\textrm{GeV}$.

\textbf{Jet reconstruction efficiency:}
\label{sec:systematics_jeff}
A small difference between the jet reconstruction efficiencies 
measured in data and simulation was observed~\cite{Aad:2011he}, 
and as this difference can affect the final measured $\mtop$ 
value, a set of pseudo experiments are performed in which jets 
from simulated events are removed at random. 
The frequency of this is chosen such that the modified jet 
reconstruction efficiency in simulation matches the value 
measured in data.
The analysis is repeated with this change and no significant 
difference is observed.

%-------------------------------------------------------------------------------
\section{Measurement of $\mtop$}
\label{sec:top measurement}
%-------------------------------------------------------------------------------
%
After applying the method described in 
Section~\ref{sec:Top-quark measurements} the top-quark mass 
is measured to be:

\begin{equation} 
  \mtop = 173.72  \pm 0.55 \stat \pm 1.01 \syst \,\textrm{GeV}.
 \label{eq:topextraction}
\end{equation}

The statistical error quoted in Eq.~(\ref{eq:topextraction}) 
is corrected for the correlation between the two $\rtt$ 
measurements of each event, as discussed in Section~\ref{sec:Top-quark measurements}.
The systematic uncertainty quoted above is the sum in quadrature 
of all the systematic uncertainties described in 
Section~\ref{sec:systematic uncertainties} and summarised 
in Table~\ref{tab:sys_all}.
Figure~\ref{fig:topresul0} shows the $\rtt$ distribution 
(left plot) with the corresponding total fit as well as its 
decomposition into signal and the multi-jet background.
The right plot in this figure shows the ellipses corresponding 
to $1$-$\sigma$ (solid line) and $2$-$\sigma$ (dashed line) 
variations in statistical uncertainty.
This measurement agrees with the previous all-hadronic 
$\mtop$ measurement performed by ATLAS in 
$7\,\textrm{TeV}$~\cite{Aad:2014zea} data, with the $\mtop$ 
measurements performed in the single-lepton and dileptonic 
decay channels~\cite{PhysRevLett.113.032002,Aad:2015nba,Khachatryan:2015hba,Aaboud:2016igd} 
and with the results of combining the Tevatron and 
LHC measurements~\cite{ATLAS:2014wva}.                         

\begin{figure}[htp]
  \centering
       {\resizebox{0.49\textwidth}{!}{%                                         
        \includegraphics{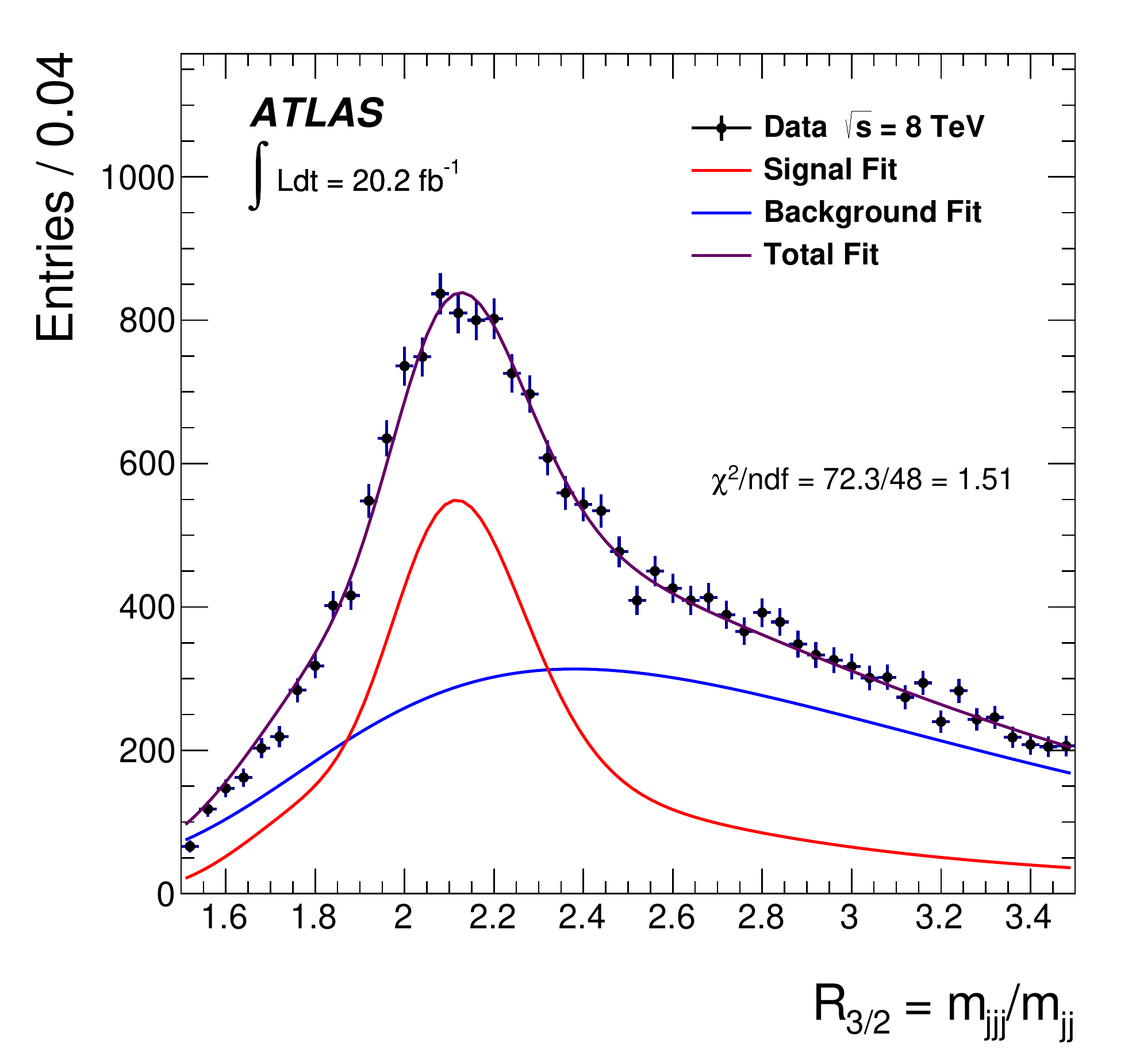}}}
        {\resizebox{0.49\textwidth}{!}{%                                         
        \includegraphics{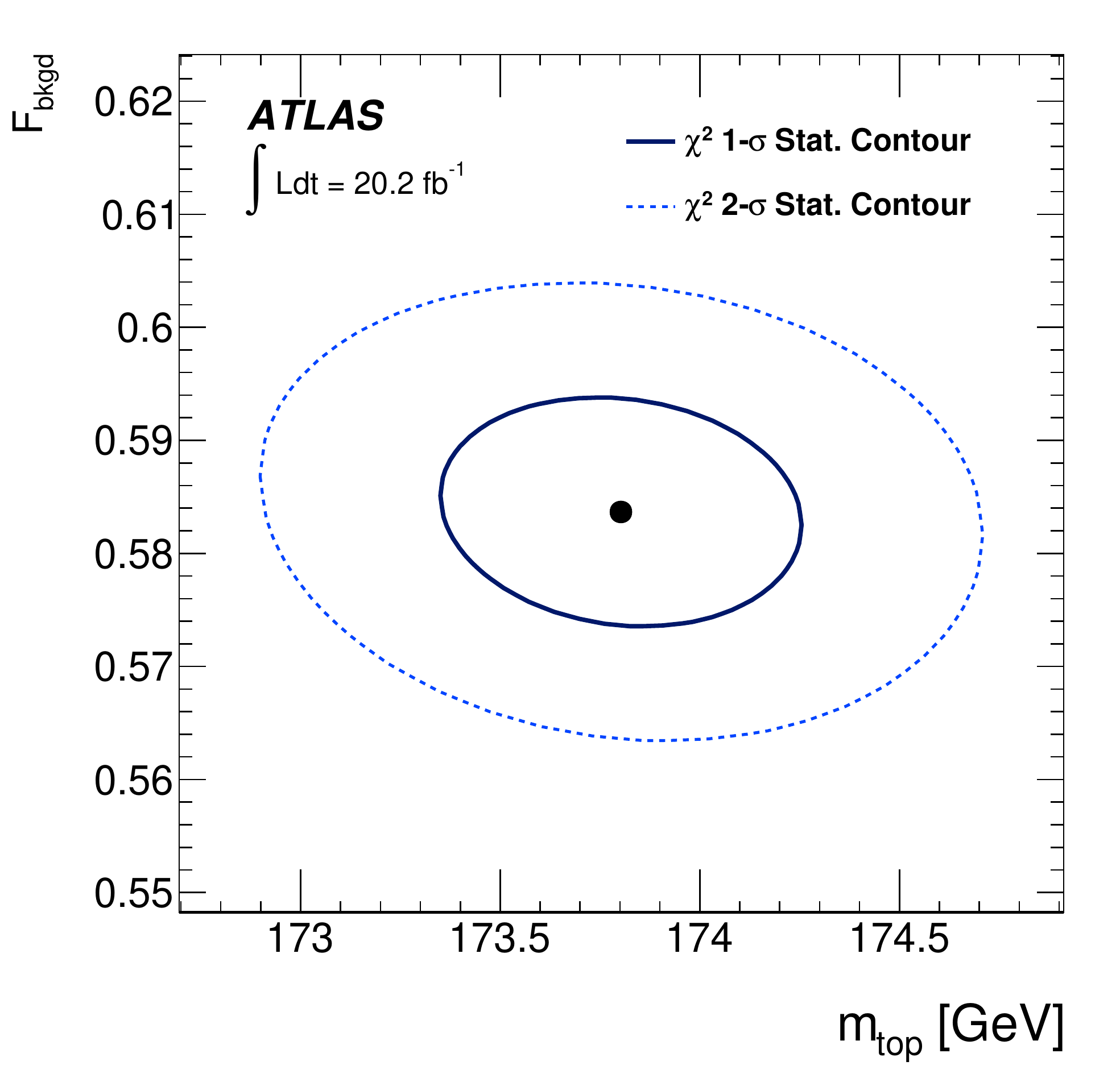}}}
  \caption{
     The left plot shows the $\rtt$ distribution in data with the  
     total fit (in magenta) and its decomposition into signal (in red) 
     and the multi-jet background (in blue). The errors shown are 
     statistical only. 
     The right plot shows the ellipses corresponding to the 
     $1$-$\sigma$ (solid line) and $2$-$\sigma$ (dashed line) 
     statistical uncertainty.
     The central point in the figure indicates the values obtained
     for $\mtop$ on the $x$--axis,
     and the fitted background fraction, $\fbkgd$, 
     obtained within the fit range of the $\rtt$ distribution
     on the $y$--axis.
     The plots do not take into account the small bias correction 
     described in Section~\ref{sec:systematics_nonclosure}.
     The top-quark mass, after this correction, 
     is $173.72 \pm 0.55 \stat \pm 1.01 \syst \,\textrm{GeV}$.
}
  \label{fig:topresul0}
\end{figure}

%-------------------------------------------------------------------------------
\section{Conclusion}
\label{sec:results}
%-------------------------------------------------------------------------------

From the analysis of $20.2\,\ifb$ of data recorded with the 
ATLAS detector at the LHC at a $pp$ centre-of-mass energy of 
$8\,\textrm{TeV}$, the top-quark mass has been measured in the 
all-hadronic decay channel of top--antitop quark pairs to be

\begin{equation} 
  \mtop = 173.72 \pm 0.55 \stat \pm 1.01 \syst \,\textrm{GeV}.
 \label{eq:topmass}
\end{equation}

This measurement is obtained from template fits to the 
$\rtt$ observable, which is chosen due to its reduced 
dependence on the jet energy scale uncertainty.
The dominant remaining sources of systematic uncertainty, 
despite the usage of the $\rtt$ observable, come 
from the jet energy scale, hadronisation modelling and 
the $b$-jet energy scale. 
This measurement agrees with the previous Tevatron and LHC 
$\mtop$ measurements, and with the results of Tevatron 
and LHC combinations.
It is about $40$\% more precise than the previous 
$\mtop$ measurement performed by ATLAS in the all-hadronic 
channel at $\rts = 7\,\textrm{TeV}$.

%-------------------------------------------------------------------------------
\section*{Acknowledgements}
\label{sec:acknowldge}
%-------------------------------------------------------------------------------

% Acknowledgements for papers with collision data
% Version 18-Jan-2017

% Standard acknowledgements start here
%----------------------------------------------
We thank CERN for the very successful operation of the LHC, as well as the
support staff from our institutions without whom ATLAS could not be
operated efficiently.

We acknowledge the support of ANPCyT, Argentina; YerPhI, Armenia; ARC, Australia; BMWFW and FWF, Austria; ANAS, Azerbaijan; SSTC, Belarus; CNPq and FAPESP, Brazil; NSERC, NRC and CFI, Canada; CERN; CONICYT, Chile; CAS, MOST and NSFC, China; COLCIENCIAS, Colombia; MSMT CR, MPO CR and VSC CR, Czech Republic; DNRF and DNSRC, Denmark; IN2P3-CNRS, CEA-DSM/IRFU, France; GNSF, Georgia; BMBF, HGF, and MPG, Germany; GSRT, Greece; RGC, Hong Kong SAR, China; ISF, I-CORE and Benoziyo Center, Israel; INFN, Italy; MEXT and JSPS, Japan; CNRST, Morocco; FOM and NWO, Netherlands; RCN, Norway; MNiSW and NCN, Poland; FCT, Portugal; MNE/IFA, Romania; MES of Russia and NRC KI, Russian Federation; JINR; MESTD, Serbia; MSSR, Slovakia; ARRS and MIZ\v{S}, Slovenia; DST/NRF, South Africa; MINECO, Spain; SRC and Wallenberg Foundation, Sweden; SERI, SNSF and Cantons of Bern and Geneva, Switzerland; MOST, Taiwan; TAEK, Turkey; STFC, United Kingdom; DOE and NSF, United States of America. In addition, individual groups and members have received support from BCKDF, the Canada Council, CANARIE, CRC, Compute Canada, FQRNT, and the Ontario Innovation Trust, Canada; EPLANET, ERC, ERDF, FP7, Horizon 2020 and Marie Sk{\l}odowska-Curie Actions, European Union; Investissements d'Avenir Labex and Idex, ANR, R{\'e}gion Auvergne and Fondation Partager le Savoir, France; DFG and AvH Foundation, Germany; Herakleitos, Thales and Aristeia programmes co-financed by EU-ESF and the Greek NSRF; BSF, GIF and Minerva, Israel; BRF, Norway; CERCA Programme Generalitat de Catalunya, Generalitat Valenciana, Spain; the Royal Society and Leverhulme Trust, United Kingdom.

The crucial computing support from all WLCG partners is acknowledged gratefully, in particular from CERN, the ATLAS Tier-1 facilities at TRIUMF (Canada), NDGF (Denmark, Norway, Sweden), CC-IN2P3 (France), KIT/GridKA (Germany), INFN-CNAF (Italy), NL-T1 (Netherlands), PIC (Spain), ASGC (Taiwan), RAL (UK) and BNL (USA), the Tier-2 facilities worldwide and large non-WLCG resource providers. Major contributors of computing resources are listed in Ref.~\cite{ATL-GEN-PUB-2016-002}.
%----------------------------------------------

%-------------------------------------------------------------------------------
% If you use biblatex and either biber or bibtex to process the bibliography
% just say \printbibliography here
\printbibliography
% If you want to use the traditional BibTeX you need to use the syntax below.
%\bibliographystyle{bibtex/bst/atlasBibStyleWoTitle}
%\bibliography{TOPQ_2015_03,bibtex/bib/ATLAS}
%-------------------------------------------------------------------------------

\newpage 
% ATLAS Collaboration author list
% Data extracted on 29-Sep-2016 for paper reference TOPQ-2015-03
%\documentclass[11pt]{article}
%\usepackage{a4wide}
%\begin{document}
\begin{flushleft}
{\Large The ATLAS Collaboration}

\bigskip

M.~Aaboud$^\textrm{\scriptsize 137d}$,
G.~Aad$^\textrm{\scriptsize 88}$,
B.~Abbott$^\textrm{\scriptsize 115}$,
J.~Abdallah$^\textrm{\scriptsize 8}$,
O.~Abdinov$^\textrm{\scriptsize 12}$,
B.~Abeloos$^\textrm{\scriptsize 119}$,
R.~Aben$^\textrm{\scriptsize 109}$,
O.S.~AbouZeid$^\textrm{\scriptsize 139}$,
N.L.~Abraham$^\textrm{\scriptsize 151}$,
H.~Abramowicz$^\textrm{\scriptsize 155}$
% ...
\bigskip
\\
% ...
$^{8}$ Department of Physics, The University of Texas at Arlington, Arlington TX, United States of America\\
% ...
$^{12}$ Institute of Physics, Azerbaijan Academy of Sciences, Baku, Azerbaijan\\
% ...
$^{88}$ CPPM, Aix-Marseille Universit{\'e} and CNRS/IN2P3, Marseille, France\\
% ...
$^{109}$ Nikhef National Institute for Subatomic Physics and University of Amsterdam, Amsterdam, Netherlands\\
% ...
$^{115}$ Homer L. Dodge Department of Physics and Astronomy, University of Oklahoma, Norman OK, United States of America\\
% ...
$^{119}$ LAL, Univ. Paris-Sud, CNRS/IN2P3, Universit{\'e} Paris-Saclay, Orsay, France\\
% ...
$^{137}$ $^{(a)}$ Facult{\'e} des Sciences Ain Chock, R{\'e}seau Universitaire de Physique des Hautes Energies - Universit{\'e} Hassan II, Casablanca; $^{(b)}$ Centre National de l'Energie des Sciences Techniques Nucleaires, Rabat; $^{(c)}$ Facult{\'e} des Sciences Semlalia, Universit{\'e} Cadi Ayyad, LPHEA-Marrakech; $^{(d)}$ Facult{\'e} des Sciences, Universit{\'e} Mohamed Premier and LPTPM, Oujda; $^{(e)}$ Facult{\'e} des sciences, Universit{\'e} Mohammed V, Rabat, Morocco\\
% ...
$^{139}$ Santa Cruz Institute for Particle Physics, University of California Santa Cruz, Santa Cruz CA, United States of America\\
% ...
$^{151}$ Department of Physics and Astronomy, University of Sussex, Brighton, United Kingdom\\
% ...
$^{155}$ Raymond and Beverly Sackler School of Physics and Astronomy, Tel Aviv University, Tel Aviv, Israel\\
% ...
$^{a}$ Also at Department of Physics, King's College London, London, United Kingdom\\
$^{b}$ Also at Institute of Physics, Azerbaijan Academy of Sciences, Baku, Azerbaijan\\
$^{c}$ Also at Novosibirsk State University, Novosibirsk, Russia\\
$^{d}$ Also at TRIUMF, Vancouver BC, Canada\\
$^{e}$ Also at Department of Physics {\&} Astronomy, University of Louisville, Louisville, KY, United States of America\\
% ...
\end{flushleft}

%\end{document}
% Created with xml2latex.py

\end{document}